\documentclass[final,3p,times,authoryear]{elsarticle}

\usepackage{amssymb}
\usepackage{epsfig}
\usepackage{centernot}
\usepackage[colorlinks,linkcolor = red, citecolor=blue,urlcolor=blue,filecolor=blue]{hyperref} 
\usepackage{url}
\usepackage{multirow}
\usepackage{graphicx}
\usepackage{enumitem}
\usepackage{lscape}
\usepackage{dsfont}
\usepackage[ruled,vlined]{algorithm2e}
\usepackage{algorithmic}
\usepackage{placeins}
\usepackage{mathtools}
\usepackage[mathscr]{euscript}
\usepackage{tikz-network}
\usepackage{slashbox} 
\usepackage{subcaption}
\usepackage{comment}

\DeclareUnicodeCharacter{0301}{\'}
\DeclareMathOperator*{\argmax}{arg\,max}

\makeatletter
\newcommand{\substackalign}[1]{%
	\vcenter{%
		\Let@ \restore@math@cr \default@tag
		\baselineskip\fontdimen10 \scriptfont\tw@
		\advance\baselineskip\fontdimen12 \scriptfont\tw@
		\lineskip\thr@@\fontdimen8 \scriptfont\thr@@
		\lineskiplimit\lineskip
		\ialign{\hfil$\m@th\scriptstyle##$&$\m@th\scriptstyle{}##$\hfil\crcr
			#1\crcr
		}%
	}%
}
\makeatother

\usepackage{amsfonts,mathrsfs,bbm, dsfont}

\newcommand{\Bf}{\mathbf{f}}

\newcommand{\Bn}{\mathbf{n}}

\newcommand{\Br}{\mathbf{r}}

\newcommand{\BL}{\mathbf{L}}
\newcommand{\BM}{\mathbf{M}}
\newcommand{\BN}{\mathbf{N}}

\newcommand{\CA}{\mathcal{A}}

\newcommand{\CC}{\mathcal{C}}

\newcommand{\CF}{\mathcal{F}}
\newcommand{\CG}{\mathcal{G}}

\newcommand{\CK}{\mathcal{K}}
\newcommand{\CL}{\mathcal{L}}

\newcommand{\CN}{\mathcal{N}}
\newcommand{\CO}{\mathcal{O}}
\newcommand{\CP}{\mathcal{P}}

\newcommand{\CT}{\mathcal{T}}

\newcommand{\CX}{\mathcal{X}}

\newcommand{\FC}{\mathfrak{C}}

\newcommand{\FL}{\mathfrak{L}}

\newcommand{\FN}{\mathfrak{N}}

\newcommand{\FS}{\mathfrak{S}}

\newcommand{\R}{\mathbb{R}} 
\newcommand{\N}{\mathbb{N}} 
\newcommand{\TRANSP}{\mathsf{T}} 
\newcommand{\indicator}{\mathds{1}}
\newcommand{\expectation}{\mathds{E}}
\newcommand{\variance}{\mathrm{Var}}

\DeclareMathOperator*{\minimize}{\mathrm{minimize}} 
\DeclareMathOperator*{\maximize}{\mathrm{maximize}} 

\newcommand{\SC}[1]{\textbf{\textcolor{red}{#1}}}
\newcommand{\CZY}[1]{\textbf{\textcolor{cyan}{#1}}}

\newtheorem{theorem}{Theorem}
\newtheorem{lemma}[theorem]{Lemma}
\newtheorem{assumption}{Assumption}
\newtheorem{definition}{Definition}
\newdefinition{remark}{Remark}
\newdefinition{example}{Example}
\newproof{proof}{Proof}
\newcommand{\proofofref}{}
\newproof{zproofof}{Proof of \proofofref}
\newenvironment{proofof}[1]
{\renewcommand{\proofofref}{#1}\zproofof}
{\endzproofof}

\begin{document}

\begin{frontmatter}


\title{Utilitarian Welfare Optimization in the Generalized Vertex Coloring Games:
	An Implication to Venue Selection in Events Planning}


\author[label1]{Zeyi Chen}
\ead{chen1417@e.ntu.edu.sg}
\affiliation[label1]{organization={School of Physical and Mathematical Sciences, Nanyang Technological University}, country={Singapore}}


\begin{abstract}
We consider a general class of multi-agent games in networks, namely the generalized vertex coloring games (G-VCGs), inspired by real-life applications of the venue selection problem in events planning. Certain utility responding to the contemporary coloring assignment will be received by each agent under some particular mechanism, who, striving to maximize his own utility, is restricted to local information thus self-organizing when choosing another color. Our focus is on maximizing some utilitarian-looking welfare objective function concerning the cumulative utilities across the network in a decentralized fashion. Firstly, we investigate on a special class of the G-VCGs, namely Identical Preference VCGs (IP-VCGs) which recovers the rudimentary work by \cite{chaudhuri2008network}. We reveal its convergence even under a completely greedy policy and completely synchronous settings, with a stochastic bound on the converging rate provided. Secondly, regarding the general G-VCGs, a greediness-preserved Metropolis-Hasting based policy is proposed for each agent to initiate with the limited information and its optimality under asynchronous settings is proved using theories from the regular perturbed Markov processes. The policy was also empirically witnessed to be robust under independently synchronous settings. Thirdly, in the spirit of ``robust coloring'', we include an expected loss term in our objective function to balance between the utilities and robustness. An optimal coloring for this robust welfare optimization would be derived through a second-stage MH-policy driven algorithm. Simulation experiments are given to showcase the efficiency of our proposed strategy.
\end{abstract}


\end{frontmatter}


\section{Introduction}
\label{sec:intro}
Selecting a right venue is crucial to the success of an event and worth emphasis in events planning. Critical criteria for being ``right''  may consider the venue's layout and size, the accessibility, technical requirements, the atmosphere or tone the planner wishes to convey, etc., and often differ across specific activities \citep{allen2022festival}. A common challenge confronted in reality is for a group of different event managers to select their preferred locations among a limited number of venues, especially when there exist mutual timetable clashes between events, and a beneficial-to-all venue assignment is always desirable. For instance, multiple clubs on a campus who intend to organize respective events in celebration of certain festival may fail to obtain their respective favorite venue due to the limited number of available function halls. A minimum requirement for any event planning is to guarantee that each event would be placed somewhere without a timetable clash with others, which can be modelled by a classical graph (vertex) coloring problem.  Vertex coloring has long been a prominent tool for modelling network problems with multifrious applications including scheduling, channel assignment, text or image segmentation, etc.; see, e.g. a survey by \cite{ahmed2012applications}. Given a connected graph (network) $G = (V, E)$ with $|V| = n$ and a collection of colors $M = \{M_1, M_2, \dots, M_m\}$ with cardinality $m$, a \textit{coloring} is a function $c: V \rightarrow M$ that assigns each vertex $V_i \in V$ a color $m_i \in M$. Denote the color assigned to a vertex $V_i$ by $c_i$. We say that a clash occurs between two vertices $V_i$ and $V_j$ when $c_i = c_j$, $e_{ij} \in E$. It is desired for many scenarios to obtain a coloring without clash, namely a \textit{proper coloring} $c^p: V \rightarrow M$ such that $c^p(V_i) \neq c^p(V_j)$ whenever $e_{ij} \in E$. Denote the space of all possible colorings by $C$ and the space of all proper colorings by $C^p$. For the sake of representation, the graph is often translated to a square \textit{location matrix} $\CL^{n \times n}$ where $\CL_{ij} = \indicator_{\{e_{ij} \in E\}}$. The \textit{nighborhood} of a vertex $V_i \in V$, $\CN(V_i)$, is defined as the set $\{V_j: \CL_{ij} = 1\}$. 

Concretely during events planning, one may represent different function halls as different colors. Let each club manager be represented by a vertex, then an edge can be linked between any pair whose event schedules clash, which establishes a connected graph (network). The goal is then to look for a \textit{proper} coloring; i.e. a coloring in which adjacent vertices are associated with distinguished colors. However, classical procedures of solving such a vertex coloring problem often provide few practical insights for event planners due to two drawbacks. One unfavorable shortcoming is that classical vertex colorings always bother a ``central agent'' to gather and manipulate with the information contained in the entire network \citep{kearns2006experimental, chaudhuri2008network}. In real-life planning, this agent with ``omniscient'' view is most times absent, and each club manager needs to propose and apply for a venue autonomously. Algorithmically speaking, though centralized colorings may converge to a proper one quickly, they require heavy workload or computation, especially when $n$ is large, which are most times out of reach in reality. The other limitation is regarding the overlooked real value, or social welfare, generated by the eventual coloring. As mentioned, managers do have varied preferences on different venues, thus beyond the properness of the eventual coloring, the satisfiability of each agent under the coloring is worth further attention.

Bearing the above concerns in mind, in this paper, we keep the agents in the network self-organized and extend the pursuit of a proper coloring to more general objectives. To eliminate any dependence on ``central agent'', each vertex is motivated to strive for some target coloring themselves through limited information sharing and mutual negotiations. In other words, instead of being ``allocated" a color, each vertex is required to ``select" a color autonomously in rounds and solve the possible improper situations themselves through compliant communications.
Such a ``self-organizing" setting better mimics the real dynamics and interactions occuring in the multi-agent system and converts the original static task to an evolutionary game, namely a Vertex Coloring Game (VCG). Meanwhile, in an attempt to include the group satisfiability into consideration, certain preference mechanism is employed to measure personal satisfiabilities. Analogue to event planners' customized preferences to different function halls, a \textit{preference function} $\phi_i: M \rightarrow \R_+$ is developed by each vertex $V_i$ to measure its preference to the current color in hold, with the total number of colors provided as a constraint. Like in many conventional game-theoretic settings, we are interested in the social welfare across the entire network which is evaluated via some \textit{welfare functional} $\Phi: C \rightarrow \R$ defined by $\Phi(c) := \Bf(\{\phi_1(c_1), \phi_2(c_2), \dots, \phi_n(c_n)\})$ which is essentially a function of all individual utilities. Such functions are often referred to as ``cardinal'' welfare functions and cover a wide range of candidates including utilitarian welfare, minmax welfare, Foster's welfare, etc. \citep{keeney1975group}. Finding a most beneficial venue assignment is thus equivalent to maximizing the social welfare without violating the constraint of a proper coloring, which is summarized in the following constrained welfare optimization problem
\begin{align}
	\label{prob: optimalcoloring1}
	\begin{split}
		\maximize_{c \in C} \quad &\Phi(c) := \Bf(\{\phi_1(c_1), \phi_2(c_2), \dots, \phi_n(c_n)\})\\
		\mathrm{subject~to} \quad & c_i \neq c_j, \forall i, \forall j \in \{j: \CL_{ij} = 1\}
	\end{split}
	\tag{$\mathsf{WO1}$}.
\end{align}

\begin{remark}
	We would like to accentuate that, the optimal coloring to \eqref{prob: optimalcoloring1}, $c^*_{WO1}$, is necessarily a \textit{Nash equilibrium}. Suppose not, since $u_i(c^*_{WO1_i}, c^*_{WO1_{-i}}) > 0$ (by the properness of $c^*_{WO1}$), the only way for $V_i$ to obtain better utility is to select another color he prefers more but different from all his neighbots' choices, otherwise his utility would be reduced to zero. This operation destines an increase in social welfare thus contracting to the fact that $c^*_{WO1}$ is optimal.
\end{remark}

Note that the set constraint of proper colorings typically do not enjoy nice geometric structures and the preference functions are completely arbitrary and purely decided by the particular event manager. Consequently, it is often difficult to explicitly control the feasibility of a new coloring when optimizing the objective in Problem~\ref{prob: optimalcoloring1}. Hence, it is natural to consider some relaxation to somehow decompose and reflect this network constraint to individual agents; i.e. each agent would commit the duty to satisfy the constraint. To achieve this goal, we define a \textit{utility function} $u_i: C \rightarrow \R_+ \cup \{0\}$ for each agent as
\begin{align}
	u_i(c) := \phi_i(c_i) \indicator_{c_i \neq c_j, \forall j \in \CL_{ij}}
\end{align}
and formulate another optimization problem
\begin{align}
	\label{prob: optimalcoloring2}
	\maximize_{c \in C} \quad &\Phi'(c) := \Bf(\{u_1(c), u_2(c), \dots, u_n(c)\})
	\tag{$\mathsf{WO2}$}.
\end{align}
Denote the optimal colorings of \eqref{prob: optimalcoloring1} and \eqref{prob: optimalcoloring2} by $c^*_{WO1}$ and $c^*_{WO2}$ in respective. In Section~\ref{sec: game formulation}, we will show that $c^*_{WO1} = c^*_{WO2}$ under several conditions.

As mentioned, the information accessible to each event manager is limited and it is extremely expensive for each event manager to collect the latest situations of all other clubs in the entire network. We integrate this consideration of information scarcity into our modelling and restrict vertices' exposures to two types of ``local'' information:
\begin{enumerate}[label=\normalfont{(Info-Type\arabic*)},leftmargin=70pt]
	\item \label{info: neighbor} A vertex can access to the information of his neighbors, including the current utility, the current color assignment and the weight in social welfare.
	\item \label{info: member} A vertex can access to the information of the members holding the same color, including the current utility, the current color assignment and the weight in social welfare.
\end{enumerate}
Despite its advantages in simulating realistic behaviours, such a VCG with restricted information may be associated with several vulnerabilities. For instance, scarce and unbalanced information can obstruct the convergence to an optimal solution to Problem~\ref{prob: optimalcoloring2}, and manager's personal incentives may inconsilient with the global interest. We illuminate three threatening features of the VCGs as follows:
\begin{enumerate}[label=\normalfont{(Feature \arabic*)},leftmargin=70pt]
	\item \label{feature: greedy} The agents are \textit{greedy}: if an agent is offered an alternative color that improves his utility, he would have a high incentive to embrace it; otherwise, when confronted with a color he disdains, he would most times reject on the assignment.
	\item \label{feature: uncomplacent} The agents are \textit{uncomplacent}: The agents are not shiftless but always active to try some other color, even if he is not aware of whether a better choice exists. 
	\item \label{feature: myopic} The agents are \textit{myopic}. Since each agent is only informed of the current situation without presaging capabilities, he can hardly realize that a temporary sacrifice on the utility may result in better personal outcomes in later rounds. 
\end{enumerate}
In each case, the optimization procedure regarding Problem~\eqref{prob: optimalcoloring2} would be affected or even undermined. For instance, the game is likely to fall stuck when any agent occupies a certain color regardless of his neighbor's benefits, or constantly rejects on taking any color he dislikes. The search space thus severely shrinks and the optimal solution may become inattainable. Fortunately, we will show in later sections that, under mild conditions and some proper strategy, the convergence to an optimal or suboptimal solution for Problem~\eqref{prob: optimalcoloring2} is immuned to the above features. 

Throughout our following discussions, the games are most modeled by discrete-time Markov Chains (MCs) with carefully designed transition matrices. This is because of the stochastic nature embedded in our game-theoretic setting, and the fact that the G-VCG obeys the Markov property: evolution of the game is memoryless and would only be determined by the present state. The corresponding state space would be the space of all possible colorings. Techniques and methods like the first-step analysis and Markov Chain Monte Carlo (MCMC) would be employed for mathematical derivations and algorithm designs.

To summarize, the main contributions of this paper are as follows. 
\begin{enumerate}[label=(\arabic*), leftmargin=2em]
	\item We extend the goal of finding a proper coloring (a feasibility problem) to finding some proper max-welfare coloring solution (a constrained optimization problem) through information restrictive VCGs as expatiated in Problem~\ref{prob: optimalcoloring1} and Problem~\ref{prob: optimalcoloring2}, which better caters for the goal for venue selection in event planning.
	\item We investigate on a special class of G-VCGs where agents are indifferent about the color assigned but only concern possible clashes, namely the Identical Preference VCGs (IP-VCGs). As a continuation to the work led by \cite{chaudhuri2008network}, we further propose and prove a stochastic upper bound of $O_p (\log n)$ for the convergence time to the optimal coloring, under mild assumptions.
	\item Novel discussions are given on broad G-VCGs where an agent's personal utility is contingent both on whether he is clash-free and the specific color assigned to him. Though a completely greedy policy no longer leads to optimality, we propose and show that a Metropolis-Hasting based policy that, while respecting agents' feature of being greedy, help the self-organizing network move towards a optimal coloring in asynchronous settings and a special class of independently synchronous settings. Theories from the regular perturbed Markov Process (RPMP) are employed in proofs.
	\item In a spirit of ``robust coloring'', we further integrate into our formulation an extra term representing the expected loss from possible complementary edge connection, to increase the robustness of the optimal coloring in the sense firstly mentioned in \cite{yanez2003robust}). To our knowledge, this is the first attempt to interact the objective of robust coloring with another optimization scheme such that the welfare and the risk are well balanced. A second-stage Metropolis-Hasting based algorithm taking the optimal solution to \ref{prob: optimalcoloring1} as input is presented to solve \ref{prob: RWO}.
	
\end{enumerate}

\section{Literature Review}
\label{sec:literature}

\subsection{Network Coloring Game for Conflict Resolving}
The concept of the self-organized VCG was informally proposed in a behavioural study reported in \cite{kearns2006experimental}, which was motivated by a self-organizing venue assignment problem among the faculty members. With the same aim of resolving conflicts (i.e. achieving some proper coloring), experiments were conducted against different network topologies and with variations to the extent of limits on information sharing among the agents. The conflict-resolving status and time were well monitored and compared across networks with distinguished features. In \cite{chaudhuri2008network}, such social interactions within a network was first formulated as a game with binary payoffs, and theoretical results were derived through ingenious combinatorial arguments. It was shown that, even the agents would adopt completely greedy strategies and are allowed to act simultaneously, a proper coloring is attainable when available colors are two more in number than the maximum degree $\Delta$ of the graph, and the procedure ends in $O(\frac{\log n}{\delta})$ rounds with probability $1 - \delta$. Almost at the same time, an independent work led by \cite*{panagopoulou2008game} studied a similar problem with a slightly different payoff scheme from a more game-theoretic perception. The authors showed that every Nash equilibrium of the corresponding VCG is feasible and locally optimal, and a characterization of the Nash equilibria was provided. 

The two pioneering works attracted numerous attention to the field of decentralized coloring game. One branch of work focuses on customized results when the VCG is restricted to certain special graph structures (see, e.g., \citep*{enemark2011does}). Another popular ramification is induced by the flexible game settings, both on game types (\citep*{pelekis2013network, carosi2019coalition}) and payoff features (\citep*{kliemann2015price}) and employed strategies (\citep*{hernandez2012distributed}). Of course, constant efforts have also been paid on improving the upper bounds for the convergence rate thus the complexity of the VCG \citep*{bermond2019long} or reduing the color supply \citep*{fryganiotis2023note}. The idea of the VCG also shed light on many engineering problem-solvings; see, e.g. \citep*{goonewardena2014minimum, marden2013overcoming, touhiduzzaman2018diversity}.

\subsection{The Use of Markov Chains in Graph Coloring}
A Markov chain is a stochastic process that models a sequence of random events in which the probability of each event depends only on the state of the system at the previous event. It has been widely used in graph coloring, which is NP-hard, due to its advantage in handling large-scale search in a systematic and efficient manner. One of the earliest Markov chain algorithms for graph coloring was proposed by \cite*{kirkpatrick1983optimization} where the transition probability depends on the difference in the number of conflicting edges between the two colorings. The algorithm proved to be effective on small to medium-sized graphs had difficulty with larger graphs. In recent years, there have been several other Markov chain algorithms proposed for graph coloring, such as the Genetic Algorithm \citep*{fleurent1996genetic, hindi2012genetic, marappan2013new}, Ant Colony Optimization \citep*{salari2005aco, dowsland2008improved}, and Particle Swarm Optimization \citep*{cui2008modified, marappan2021solving}. Promising results have been shown on a wide range of graph coloring problems.

Markov chains have also been applied for color samplings. Aiming to approximately counting the number of the k-colorings in a graph, \cite{jerrum1995very} converted this counting problem to estimating the mixing time of a Markov Chain. Inspired by the term of Glauber Dynamic in statistical physics, he presented an approach to ramdomly sampling colorings with maximum degree $\Delta$ in $O(n \log n)$ time with at least $2\Delta + 1$ colors provided. \cite{vigoda2000improved} later proved that it suffices for the number of colors to be only $\frac{11}{6} \Delta$ for the $O(n \log n)$ bound. Further results were developed for some graphs with special features (see, e.g., \citep*{hayes2006coupling, hayes2007randomly}). Recently, Vigoda's result on general graphs was improved by \cite*{chen2019improved} who proved that the chains are rapidly mixing in $O(n \log n)$ time when there are at least $(\frac{11}{6} - \epsilon_0) \Delta$ colors where $\epsilon_0$ is a small positive constant, using the linear programming approach.

\subsection{The Metropolis-Hasting Algorithm for Optimization}
\label{ssec: MH algo}
The Metropolis-Hastings algorithm is a versatile algorithm for solving complex optimization problems that was first introduced in a seminal paper by \cite{metropolis1953equation} and further modified by \cite{hastings1970monte} who included a correction factor to ensure detailed balance. It relies on constructing a Markov chain that has the desired target distribution as its stationary distribution, from which it generates a sequence of samples. The algorithm has since been adapted for engineering applications emerging in various fields, including but not limited to signal processing \citep*{luengo2013fully, vu2014particle, marnissi2020majorize}, self-reconfiguration systems \citep*{pickem2015game}, task allocation \citep*{hamza2021metropolis, moayedikia2020optimizing}, etc.

A snapshot of the algorithm is as follows: In a finite-state space $S$, denote the target distribution by $\pi(X)$ where $X$ is the variable ot interest. As an initialization one constructs a starting state $x_0$ and an irreducible proposal transition matrix $Q(x'|x)$ to generate a new state $x'$ from $x$. The transition probability is then multiplied by an acceptance rate 
\begin{align*}
	\alpha(x \rightarrow x') = \min \{1, \frac{\pi(x')Q(x | x')}{\pi(x)Q(x'|x)}\} 
\end{align*}
and the updated transition probability, namely the target transition probability, is 
\begin{align*}
	P(x'|x) &= Q(x'|x)\alpha(x \rightarrow x'), \quad x' \neq x;\\
	P(x|x) & = 1 - \sum_{x' \neq x} P(x'|x).
\end{align*}
Note that the target transition probability satisfies the local balanced equation
\begin{align*}
	\pi(x) P(x'|x) = \pi(x') P(x | x')
\end{align*}
thus the target distribution o$\pi(X)$ is exactly a stationary distribution of the Markov process induced by $P$. The uniqueness of the stationary distribution is given by irreducibility of the chain. 

\section{Game Formulation}
\label{sec: game formulation}
\vspace{10pt}
\subsection{Setting}
\label{subsec: setting}
Let us first introduce the notations and terminologies that are used throughout this paper. Notations of the graph structure and the color collection follow from Section~\ref{sec:intro}, and the maximum degree of the graph $G$ is denoted by $\delta(G)$. We require $m \geq \CX(G)$ where $\CX(G)$ is the \textit{chromatic number} of G; i.e. the least number of colors that enables a proper coloring. It is denoted by $V_j \rightarrow V_i$ if the information of $V_j$ can be shared to $V_i$, and $V_i$'s \textit{family} is defined as the set $\CF_i:= \{V_j: V_j \rightarrow V_i\}$. Let $c^k \in C$ (k = 1, 2, \dots, $|C|$) be the coloring realizations. By adding a subscript $i \in \{1, 2, \dots, n\}$ we specify the particular color assigned to $V_i$; e.g. $c^k_i$ specifies the assignment under the coloring $c^k$. When there is a change in coloring, say from $c^k$ to $c^l$, let $\Delta u_i (c^k \rightarrow c^l):= u_i(c^l) - u_i(c^k)$ and $\Delta \Phi (c^k \rightarrow c^l) := \Phi (c^l) - \Phi (c^k)$ denote the difference in $V_i$'s utility and the welfare respectively under the two colorings. To emphasize on the color choice of a particular agent, we sometimes specify $u_i(c)$ as $u_i(c_i, c_{-i})$. 

We say an agent is \textit{active} in round $t$ if he wishes to update his color choice. Starting from any coloring $c(0) \in C$, the game continues in discrete rounds. Let $c_i(t)$ denote the color assigned to $V_i$ in round $t$ and $\CA(t)$ denote the \textit{active set} containing all active agents in round $t$. In each round, every active agent, driven by either \ref{feature: greedy} or \ref{feature: uncomplacent}, would attempt to update his color from certain available color set (or stategy set) $\CC_i$ under some policy, while the inactive ones keep their colors unchanged. This induces a discrete-time Markov Chain on the coloring space $C$ with its transition matrix determined by the policy. A formal definition of the \textit{Generalized Vertex Coloring Game} (G-VCG) is given as follows:
\begin{definition}[G-VCG]
	\label{def: G-VCG}
	A G-VCG is a quadruplet $\CG = (V, G, \CC, u)$ where:
	\begin{itemize}[label*=--]
		\item $V = \{V_1, V_2, \dots, V_n\}$: the set of agents (vertices).
		\item $G(V, E)$: the connected network constraining agents' information, which is apriori to agents.
		\item $\CC_i \ni c_i$: the set of available colors (pure strategies) for $V_i$.
		\item $\CC = \prod_i \CC_i$: the set of coloring profiles.
		\item $u = (u_1, u_2, \dots, u_n)$: the set of utility functions.
	\end{itemize}
\end{definition}

In the following sections, there will be discussions and experiments of G-VCGs on both \textit{asynchronous} and \textit{synchronous} conditions. A discrete game is said to be asynchronous if at most one agent is active in each round, while synchrony allows multiple active agents to update in the same round. In this paper, two particular types of synchrony, namely independent synchrony and complete synchrony, will be discussed.

\begin{definition}[Independent and completely synchronous G-VCGs]
	A G-VCG $\CG$ is said to be \textit{independently synchronous} if in each round $t$ of a G-VCG $\CG$, the probability that $V_i$ becomes active, denoted by $\omega_t \in (0, 1]$, is identical across $\forall V_i \in V$. A \textit{completely synchronous} game has all agents be active in each round; i.e. $\omega_t = 1, \forall t \in \N$.
\end{definition}

\subsection{Essential Assumptions}
\label{subsec: assumptions}
In this part of the section, we make several essential assumptions, most of which are to be assumed thoughout the paper unless mentioned otherwise.

Focusing on maximizing the total utility of the network with little regard to the utility distribution, we now make an additional assumption that the welfare function is \textit{utilitarian} or \textit{weighted utilitarian} where the weights are priori information to the agents. In later sections we will see that such welfare formulas would enable individual agents to evaluate their contributions to the total welfare by selecting a new color, even without refering to the entire network.
\begin{assumption}[Utilitarian/Weighted-Utilitarian Welfare]
	The welfare function $\Phi: C \rightarrow \R$ is defined by 
	\label{assp: utilitarian welfare}
	\begin{align*}
		\Phi(c) = \sum_{V_i \in V} w_i u_i(c)
	\end{align*}
	where $w_i \geq 0$ and $\sum_i w_i = 1$.
\end{assumption}

As mentioned in Section~\ref{sec:intro}, our main concern is whether the optimal solutions of \eqref{prob: optimalcoloring2} coincide with the ones of \eqref{prob: optimalcoloring1}. It is straitforward that if $c^*_{WO2}$ is a proper coloring, then $c^*_{WO1} = c^*_{WO2}$; otherwise $c^*_{WO1}$ would be a better solution for \eqref{prob: optimalcoloring2}. Yet, whether $c^*_{WO2}$ is proper highly depends on the abundance of colors provided. We make another assumption on the relationship between the caridinality of the color set and the maximum degree of the graph.
\begin{assumption}[Abundant Colors]
	\label{assp: colors}
	The number of colors provided is at least one more than the maximum degree of the graph; i.e.
	\begin{align*}
		|M| \geq \delta + 1.
	\end{align*}
\end{assumption}
Assumption~\ref{assp: colors} guarantees $c^*_{WO2}$ to be proper because each vertex $V_i$ always has a color that is distinguished from the neighborhood; i.e. 
\begin{align}
	\forall c_{-i} \in \prod_{j \neq i} \CC_j,~ \exists c_i \in \CC_i \text{ such that } u_i(c_i, c_{-i}) > 0.
	\label{optimum coincide}
\end{align}

To show that Assumption~\ref{assp: colors} is the weakest to ensure the properness of $C^*_{WO2}$ in whatever graph structures, we would like to give a counterexample when cases with $|M| = \delta$ thus $|M| \geq \delta$ may fail.

\begin{example}[$|M| \geq \delta$ is insufficient]
	\label{exp: insufficient color}
	\vspace{10pt}
	Consider the network structure in Figure~\ref{fig: countereg0} where each vertex has equal weight in contribution to the social welfare. Suppose $M = (Red, Green, Blue)$ so that $|M| = \delta = 3$. The preferences of each vertex on each color are given in Table~\ref{table: preference}. One can easily observe that the optimal coloring $c^*_{WO2}$ is as in Figure~\ref{fig: countereg1} where $V_1$, $V_2$ and $V_4$ all obtain their top-preferred colors, and the optimal welfare is 
	\begin{align*}
		\Phi(c^*_{WO2}) = \frac{1}{4} (0 \times 1 + 0 \times 1 + 1 \times 10 + 1 \times10) = 5.
	\end{align*}
	In this case, $V_1$ and $V_2$ compromises to bear in a clash whatever color he selects, therefore $c^*_{WO2}$ is not proper. 
	\begin{figure}[!htb]
		\begin{minipage}{0.48\textwidth}
			\centering
			\begin{tikzpicture}
				\Vertex [label = $V_1$] {V1}  \Vertex [x = 2, label = $V_2$]{V2}
				\Vertex [x = 2, y = 2 , label = $V_3$]{V3} \Vertex [x = 4, y = 0, label = $V_4$]{V4}
				\Edge (V1)(V2)
				\Edge (V2)(V3)
				\Edge (V2)(V4)
				\Edge [bend=-45](V1)(V4)
				\Edge [bend = 45](V1)(V3)
			\end{tikzpicture}
			\caption{The network corresponding to Example~\ref{exp: insufficient color} where $|M| = \delta$ fails}
			\label{fig: countereg0}
		\end{minipage}
		\hfill
		\begin{minipage}{0.48\textwidth}
			\centering
			\begin{tikzpicture}
				\Vertex [label = $V_1$, color = red] {V1} 
				\Vertex [x = 2, label = $V_2$, color = red]{V2}
				\Vertex [x = 2, y = 2 , label = $V_3$, color = green]{V3} 
				\Vertex [x = 4, y = 0, label = $V_4$, color = blue]{V4}
				\Edge (V1)(V2)
				\Edge (V2)(V3)
				\Edge (V2)(V4)
				\Edge [bend=-45](V1)(V4)
				\Edge [bend = 45](V1)(V3)
			\end{tikzpicture}
			\caption{The optimal coloring $c^*_{WO2}$ for Example~\ref{exp: insufficient color}}
			\label{fig: countereg1}
		\end{minipage}
	\end{figure}
	
	\begin{table}[h!]
		\centering
		\begin{tabular}{|l|*{4}{c|}}
			\hline
			\backslashbox{Color}{Vertex}
			&\makebox[3em]{$V_1$}&\makebox[3em]{$V_2$}&\makebox[3em]{$V_3$}
			&\makebox[3em]{$V_4$}\\\hline
			Red &1&1&1&1\\\hline
			Green &1&1&10&1\\\hline
			Blue &1&1&1&10\\\hline
		\end{tabular}
		\caption{Preferences on colors for Example~\ref{exp: insufficient color}}
		\label{table: preference}
	\end{table}
	\end{example}
	
Moreover, special attention should be given to the step when the active agents modify their color choices. If not well regulated, the game would easily become vulnerable and two typical jeopardies include:
\begin{enumerate}[label=(\roman*)]
	\item \textit{Small search space}: Due to  \ref{feature: greedy}, every agent would incline to top-preferred colors and is loath to adopt other ones. This would severely reduce the search space of the process thus the optimal coloring is often attainable.
	\item \textit{Deadlock}: Due to \ref{feature: myopic} and \ref{feature: greedy}, an agent is not aware that sacrifice would bring ``win-win" outcomes for both himself and the neighbors later. This would result in deadlocks in network, especially when the preference systems differ across the neighbors, when imprudent agents pursues temporary self-seeking colors while impairing the neighborhood's progress.
\end{enumerate}
In reality, one common approach to avoid these emotional behaviors is to involve in randomness (e.g. by lottery). We make an analogical assumption here yet still respect agents' acceptance or rejection on the new color.
\begin{assumption}[Random Color Update with Policy Acceptance]
	We assume the following rules during the color updates of active agents in each round:
	\begin{enumerate}[label=\normalfont{(R\arabic*)},leftmargin=30pt]
		\item \label{assp: random1}An active agent, say $V_i$, would first randomly select a color $c_{new}$ from a color set $\CC_i$. Let $\CC_i = M$ in each round unless mentioned otherwise.
		\item \label{assp: random2}After given $c_{new}$, $V_i$ is given a transient probation period, during which he is only aware of $u_i(c_{new}, c_{-i}(t))$ and possible influences to his family members. Utilities under other unchosen colors are not accessible.
		\item \label{assp: random3} Based on the possible utility, a decision of acceptance (update the new color) or rejection (keep the old color) would be made by $V_i$ under some policy. 
	\end{enumerate}
	\label{assp: random}
\end{assumption}
When making a decision, the latest information that an active agent can refer to is the current situations of the family members (which are obtained in the last round); i.e. the active agent $V_i$ assumes $c_{-i}(t + 1) = c_{-i}(t)$ when deciding whether to accept the offer. The completely greedy policy is defined as follows:
\begin{definition}[Completely Greedy Policy]
	A completely greedy policy accepts whatever colors leading to better utility and rejects any color implying a worse or identically bad transcient utility; i.e. given $V_i$ active in round $t$ and a new color $c_{new}$,
	\begin{align}
		c_i(t + 1) = \begin{cases}
			c_{new} & \textit{if } u_i(c_{new}, c_{-i}(t))  > u_i(c_i(t), c_{-i}(t))\\
			c_i(t) & \textit{if } u_i(c_{new}, c_{-i}(t))  \leq u_i(c_i(t), c_{-i}(t)).
		\end{cases}
		\label{eqn: completely_greedy}
		\tag{CGP}
	\end{align}
\end{definition}

Here is a brief summary on the game procedure:
\begin{enumerate}[label=(\roman*)]
	\item Before round 0, an initial coloring $c(0)$ is assigned to the network $G$. Each agent $V_i$ calculates his utility by evaluating $c_i(0)$ and detect clashes in his family.
	\item In round $t \geq 0$, each element of $\CA(t)$ pursues a new color by randomly sampling a color $j \in M$. The inactive agents remain their colors in the previous rounds; i.e. $c_l(t + 1) = c_l(t)$ for $V_l \notin \CA(t)$.
	\item Each $V_k \in \CA(t)$ decides whether to accept $j$ under certain acceptance policy. If $j$ is accepted, $c_k(t + 1) = j$; otherwise $c_k(t + 1) = c_k(t)$. The round t is finished and the game steps to round $t + 1$.
\end{enumerate}

\section{Identical Preference VCGs}
\label{sec: IP-VCG}
In this section, we focus on a particular category of the G-VCGs called Identical Preference VCGs (IP-VCGs), whose definition is given as below.
\begin{definition}[IP-VCGs]
	A G-VCG is an IP-VCG if each vertex lays identical preference on every color; i.e.
	\begin{align*}
		\phi_i(m) = \phi_i,~ \forall m \in M
	\end{align*}
	where $\phi_i$ is a constant. In other words, for each $V_i \in V$, his utility function $u_i$ is indifferent about the specific color assigned. In reality, such situations are frequently witnessed when event managers are in a rush such that their goal is to confirm a feasible venue regardless of other preferences.
\end{definition}

Note that the social welfare of an IP-VCG is 
\begin{align*}
	\begin{split}
		\Phi(c) &= \sum_{i = 1}^n w_i \phi_i(c_i) \indicator_{c_i \neq c_j, \forall j \in \CL_{ij}}\\
		& \leq \sum_{i = 1}^n w_i \phi_i
	\end{split}
\end{align*}
where the equality holds if and only if $\indicator_{c_i \neq c_j, \forall j \in \CL_{ij}} = 1$ for $\forall i \in \{1, 2, \dots, n\}$. Therefore, solving Problem~\eqref{prob: optimalcoloring2} for IP-VCGs is equivalent to finding a proper coloring which recovers the binary-payoff setting in previous works (see, e.g. \cite{kearns2006experimental, chaudhuri2008network}). Thus, in the rest of this section, we restrict our attention to the simplified problem with utility
\begin{align}
	\label{eqn: binary utility}
	u_i(c) = \indicator_{c_i \neq c_j, \forall j \in \CL_{ij}}, i = \{1, 2, \dots, n\}.
\end{align}

\subsection{Convergence of Completely Synchronous IP-VCGs to Optimality under the CGP}
\label{subsec: convergence IP-VCG}
We examine in this subsection whether the convergence of an IP-VCG to its optimal coloring under \ref{eqn: completely_greedy} which is often driven by \ref{feature: greedy}. As mentioned, it suffices to answer this question: can a proper coloring be achieved in a game with binary utilities~\eqref{eqn: binary utility} under \ref{eqn: completely_greedy}? It is obvious that an active agent with zero utility would accept any color unused by his neighbor to avoid clashes, and those agents with utility one, though can still be active, would never accept any new color because they have already attained their best personal utilities. Under our Assumption~\ref{assp: colors}, it is straightforward that asynchronous IP-VCGs will converge to an optimal coloring in finite rounds since there are always extra colors different from all neighbors' for each agent to select, and the probability of such a selection is positive. In general, analysis on an arbitrary synchronous G-VCG could be quixotic for tremendous randomness and the optimality may not be guaranteed. However, we observe that the extreme setting of complete synchrony brings peculiar properties which are worth an extra attention. As such, in the rest of the section, we concentrate our arguments on the completely synchronous cases.

We would like to point out that a similar problem was discussed in the seminal work \cite{chaudhuri2008network} which also concerned about the binary utility VCG. The main difference of their setting is that, unlike in Assumption~\ref{assp: random} where we assume $\CC_i = M$, they restricted the available color set for each agent to be the colors currently unused by his neighbors; i.e. $\CC_i(t) = M \backslash \{c_j(t): V_j \in \CN(V_i)\}$. Though could be more efficient, their setting suffers from a severely reduced search space because state transfers are often only allowed in one direction, which impairs possible generalizations. Besides, their setting requires a stronger assumption than our Assumption~\ref{assp: colors} that the number of colors must be at least two more than the maximum degree. Notice that the counterexample they gave on a failure of convergence when $m = \delta(G) + 1$, however, can be solved in our setting for better color availability. We place their example here for the sake of completeness and better understanding.

\begin{example}[Theorem 2 in \cite{chaudhuri2008network}]
	Consider $G$ being a cycle with five vertices $V = \{v_1, v_2, v_3, v_4, v_5\}$ thus $\delta(G) =  2$. Given 3 colors, say $R$, $G$ and $B$, and suppose that the initial configuration is $(R, G, B, B, G)$. In the scenario when all agents with zero utility are active and obtain colors unused by neighbors, $v_3$ and $v_4$ will always be active yet keep clashing with each other, thus a proper coloring would never be reached. Nevertheless, under our setting and Assumption~\ref{assp: random}, this conflict can be easily resolved; e.g.  when $v_3$ selects the $R$ while $v_4$ does not, the configuration becomes $(R, G, R, B, G)$. 
\end{example}

Before stepping further, we first introduce a special type of Markov Chain, namely Absorbing Markov Chain (AMC), which will be our main tool of proving the convergence. 
\begin{definition}[Absorbing Markov Chain] 
	Given a Markov Chain $(Z_t)_{t \in \mathbb{N}}$, a state $i$ is called absorbing if $ \Pr[Z_{t + k} = i \mid Z_t = i] = 1 $, $\forall k \in \mathbb{N}$, otherwise is called transient. The chain $(Z_t)_{t \in \mathbb{N}}$ is defined to be an absorbing Markov Chain (AMC) if there exists at least one absorbing state that is accessible from any transient state.
	\label{def:AMC}
\end{definition}

In the context of a 0-1 VCG, which is a simplification of IP-VCGs, one may use a list with binary digits to represent the utility of players after each round. Such utility lists have a length of $n$ with each element representing the utility of a vertex, and there are $2^n$ possible outcomes. Additionally, since it is impossible for all but one players to have utility equal to 1, the total number of possible cases is reduced by $n$. Let $S$ denote the utility space consisting of these $2^n - n$ utility lists on which a Markov Chain $(X_t)_{t \in \N}$ can be run. The transition probability reflects the likelihood of a change in utility. The following Lemma~\ref{lemma:AMC} states that $(X_t)_{t \in \N}$ is an AMC.
\begin{lemma}[AMC on Utility Space]
	Let Assumptions~\ref{assp: utilitarian welfare}, \ref{assp: colors} and \ref{assp: random} hold. Consider a completely synchronous IP-VCG under \ref{eqn: completely_greedy} with utility functions \ref{eqn: binary utility}. Given the finite utility space $S$ containing all binary utility lists $\{L^1, L^2, \dots, L^{2^n - n}\}$, a discrete-time Markov Chain $(X_t)_{t \in \N}$ is established with transition matrix $P$ whose elements are defined as $P_{kl} = \Pr[X_{t + 1} = L^l | X_{t} = L^k]$. Then $(X_t)_{t \in \N}$ is an AMC.
	\label{lemma:AMC}
\end{lemma}
\begin{proofof}{Lemma~\ref{lemma:AMC}}
See Appendix~\ref{sapx: convergence IP-VCG}.
\end{proofof}

\begin{remark}
	\label{rmk: low to high}
	Note that $(X_t)_{t \in \N}$ always jumps from a low-welfare state to a high-welfare state but never in the reverse direction, because agents with utility one would never accept a new color thus remains his utility forever under the completely greedy policy.
\end{remark}

A classical property of an AMC is that it will be absorbed eventually, which means the convergence to a proper coloring in our case.
\begin{theorem}[Convergence to Optimality]
	Let Assumptions~\ref{assp: utilitarian welfare}, \ref{assp: colors} and \ref{assp: random} hold. Any completely synchronous IP-VCG under \ref{eqn: completely_greedy} converges to an optimal coloring $c^*_{WO2}$.
	\label{thm: IP-VCG convergence}
\end{theorem}
\begin{proofof}{Theorem~\ref{thm: IP-VCG convergence}}
See Appendix~\ref{sapx: convergence IP-VCG}.
\end{proofof}

\subsection{Stochastic Boundedness on the Time to Convergence of Completely Synchronous IP-VCGs}
\label{subsec: stochastic boundedness}

Given an optimal convergence of IP-VCGs, we are interested in the time to convergence with complexity measurements. In their setting of $\CC_i(t) = M \backslash \{c_j(t): V_j \in \CN(V_i)\}$, \cite{chaudhuri2008network} proposed and proved that the game ends in $O(\log \frac{n}{\epsilon})$ rounds with probability at least $1 - \epsilon$ when $m \geq \delta + 2$. They took advantage of an important property that any zero-utility agent could earn a utility rise in two consecutive rounds with a probability above some positive constant. It supprisingly 
proves that this property extends to the completely synchronous 0-1 IP-VCGs thus to all IP-VCGs under \ref{eqn: completely_greedy}. This is given as Theorem~\ref{thm: const_prob_2rounds} below and the proof is an anlogue to Lemma 3 in \cite{chaudhuri2008network} with subtle adjustments. 
\begin{theorem}[Individual Utility Rise in Two Consecutive Rounds]
	Let Assumptions~\ref{assp: utilitarian welfare}, \ref{assp: colors} and \ref{assp: random} hold. Consider a completely synchronous IP-VCG under \ref{eqn: completely_greedy} with utility functions \ref{eqn: binary utility}. Suppose $V_i \in V$ has a clash in round $t$ thus $u_i(t) = 0$, then with at least a constant probability, he would become clash-free and obtain a positive utility after two rounds; i.e.
	\begin{align*}
		\Pr [u_i(t + 2) > 0 | u_i(t) = 0] \geq \BL, \quad \forall t \in \N
	\end{align*}
	where $\BL$ is a positive constant number.
	\label{thm: const_prob_2rounds}
\end{theorem}

\begin{proofof}{Theorem~\ref{thm: const_prob_2rounds}}
See Appendix~\ref{sapx: stochastic boundedness}.
\end{proofof}

The constant lower bound, though relatively small, is useful because it provides some deterministic information in a stochastic self-organizing game scenario. More importantly in our discussions, the constant probability for a utility rise in two consecutive rounds sheds light on the behavior of the Markov Chain as defined in Theorem~\ref{lemma:AMC}, besides the upshot of being absorbed. The following lemma demonstrates respective bounds on the expectation and variance of the number of rounds to convergence. 

\begin{lemma}[Bounded Expectation and Variance for Time to Convergence]
	Let Assumptions~\ref{assp: utilitarian welfare}, \ref{assp: colors} and \ref{assp: random} hold. Consider a completely synchronous IP-VCG $\CG$ under \ref{eqn: completely_greedy} associated with a Markov Chain $(X_t)_{t \in \N}$ as defined in Theorem~\ref{lemma:AMC}. Then $\CG$ is \textit{expected} to converge to an optimal coloring to \eqref{prob: optimalcoloring2}, $c^*_{WO2}$, in $\CO \log n$ rounds; in addition, the variance of the time to convergence is $\CO (\log n)^2$.
	\label{lemma: expectation-variance}
\end{lemma}

\begin{proofof}{Lemma~\ref{lemma: expectation-variance}}
See Appendix~\ref{sapx: stochastic boundedness}. Here is a sketch for our proof: Again we focus on the representative candidate of IP-VCGs with binary utility functions \ref{eqn: binary utility}. To most utilize the information given in Theorem~\ref{thm: const_prob_2rounds}, we generate $(X_t)_{t \in \N}$ to jump two steps at a time and then double the number of steps. We also pay special attention to the first migration of $(X_t)_{t \in \N}$ and conduct the so-called ``first-step analysis'' to derive out certain induction relationship on the expected step between different initial states.
\end{proofof}

We are now ready to state the main theorem in this section, in which we show the stochastic boundedness of $T$, the time (i.e. number of steps) to convergence.
\begin{theorem}[Stochastic Boundedness of the Time to Convergence]
	Let Assumptions~\ref{assp: utilitarian welfare}, \ref{assp: colors} and \ref{assp: random} hold. Consider a completely synchronous IP-VCG $\CG(V, G, \CC, u)$ under \ref{eqn: completely_greedy} associated with a Markov Chain $(X_t)_{t \in \N}$ as defined in Theorem~\ref{lemma:AMC}. Let $T^{(n)}$ denote the number of steps needed for $\CG$ to converge in an optimal coloring $c^*_{WO2}$. Then $T^{(n)}$ is $\CO_p (\log n)$; i.e. $\frac{T^{(n)}}{\log n}$ is bounded in probability as
	\begin{align*}
		\forall \epsilon > 0, \exists \BM_\epsilon \in \R_+, \BN_\epsilon \in \N_+ \rightarrow \Pr[|\frac{T^{(n)}}{\log n}| > \BM_\epsilon] < \epsilon, \forall n > N_\epsilon.
	\end{align*}
	Moreover, $\BN_\epsilon$ can be taken as a constant independent of $\epsilon$ when $n$ is fixed.
	\label{thm:stochastic}
\end{theorem}
\begin{proofof}{Theorem~\ref{thm:stochastic}}
See Appendix \ref{sapx: stochastic boundedness}.
\end{proofof}

\section{A Metropolis-Hasting Based Optimal Policy for G-VCGs}
\label{sec: MH-policy}
In this section, we wish to examine the likelihood for the self-organizing G-VCGs in more general settings to attain its optimal coloring, possibly with some carefully-designed acceptance policy. A rudimentary scrutiny would reveal that the completely greedy policy \ref{eqn: completely_greedy} would no longer guarantee a convergence to optimality as a consequence of \ref{feature: greedy} and \ref{feature: myopic}. Below is a simple example.
\begin{example}
	\label{exp: CGP_fails}
	Consider the graph in Figure~\ref{graph:CGP_fails} and the color preference matrix in Table~\ref{table: CGP_fails}. Assumption~\ref{assp: colors} would be satisfied when three colors are provided, namely $R, G, B$. Given the initial color assignment $c(0): = (c_1(0), c_2(0), c_3(0))$ to be $(R, G, B)$. The optimal coloring would be $(G, B, G)$. However, under \ref{eqn: completely_greedy}, $V_2$ would never accept another color because $u_2(G, c_{-2}) > u_2(c_2, c_{-2})$ for $\forall c_2 \neq G$. Consequently, $V_1$ and $V_3$ would never accept $G$ when active because, for example, $u_1(c_1, G, c_3) > u_1(G, G, c_3)$ for $\forall c_1 \neq G$. The game thus never converges to its optimality.
	
	\begin{figure}[!ht]
		\centering
		\begin{tikzpicture}
			\Vertex [label = $V_1$] {V1}  \Vertex [x = 2, label = $V_2$]{V2}
			\Vertex [x = 4, label = $V_3$]{V3} 
			\Edge (V1)(V2)
			\Edge (V2)(V3)
		\end{tikzpicture}
		\caption{The graph in Example~\ref{exp: CGP_fails}}
		\label{graph:CGP_fails}
	\end{figure}
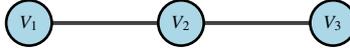
	\begin{table}[h!]
		\centering
		\begin{tabular}{|l|*{3}{c|}}
			\hline
			\backslashbox{Color}{Vertex}
			&\makebox[3em]{$V_1$}&\makebox[3em]{$V_2$}&\makebox[3em]{$V_3$}\\\hline
			Red &1&1&1\\\hline
			Green &10&10&10\\\hline
			Blue &1&2&1\\\hline
		\end{tabular}
		\caption{The preference matrix in Example~\ref{exp: CGP_fails}}
		\label{table: CGP_fails}
	\end{table}
\end{example}
Therefore, it is necessary for the agent temporarily with high utility to be a bit more ``altruistic'' to avoid such standstills.

Among literatures on solving the agent-based evolutionary optimization procedure, a similar network setting was witnessed in \cite*{marden2012revisiting} where a policy based on ``log-linear learning'' was evaluated. Nevertheless, this approach does not apply to G-VCGs because they assumed
\begin{enumerate}[label=(\roman*)]
	\item The game is \textit{potential}; i.e. $\Delta u_i = \Delta \Phi$. This is not true in our setting as a modification on an agent's choice would bring not only a difference to his utility but also to his neighbors'.
	\item For any active agent, the utility for all choices must be exhausted to make a decision. This violates \ref{assp: random2} in our Assumption~\ref{assp: random} and more space is required to store the latest utilities.
\end{enumerate}

Alternatively, inspired by the Metropolis-Hasting algorithm as reviewed in \ref{ssec: MH algo}, we propose another delicate policy adpative to our setting of G-VCGs with sound converging behaviours. We will call it \textit{MH-Policy} and details are given in Section~\ref{subsec: MH-policy}. We would like to mention that the spirit has been informally applied to task allocation of robots in \cite{hamza2021metropolis}. Yet, there lacked a mathematical discussion, from agents' standpoints, on whether the policy violates agents' incentives (e.g. \ref{feature: greedy} and \ref{feature: myopic}); otherwise the self-organizing agents woudl be reluctant to follow the rule. The gap is then filled in this section. Note that instead of the explicit stationary distribution, our focus is on the support of its stationary distribution, namely the \textit{stochastically stable states}.

\subsection{Detour: Regular Perturbed Markov Processes and Resistance Tree}
\label{subsec: detour}
It is acknowledged that, despite the mentioned limitations, we still gain parts of our insights from the work led by \cite{marden2012revisiting}. Specifically, this detour section introduces some preliminary terminologies and results from \cite*{young1993evolution} that will be invoked in the proofs of our later results. We first give the definition of a regular perturbed Markov process (RPMP).
\begin{definition}[Regular Perturbed Markov Process]
	Let $(Z_t^0)_{t \in \N}$ be a finite-state Markov chain over the finite-state space $S$ with a transition matrix $P^0$. We refer to it as an \textit{unperturbed} process. Consider a \textit{perturbed} process $(Z_t^\epsilon)_{t \in \N}$ such that the extent of perturbation can be indexed by a scalar $\epsilon > 0$, and let $P^\epsilon$ be the associated transition matrix. Then $(Z_t^\epsilon)_{t \in \N}$ is defined to be a \textit{regular perturbed Markov process} (RPMP) if the following conditions are satisfied:
	\begin{enumerate}[label=(\roman*)]
		\item \label{def: RPMC1}$(Z_t^\epsilon)_{t \in \N}$ is irreducible, $\forall \epsilon > 0$.
		\item \label{def: RPMC2}$\lim_{\epsilon \rightarrow 0^+} P_{s \rightarrow s'}^\epsilon = P_{s \rightarrow s'}^0$, $\forall s, s' \in S$.
		\item \label{def: RPMC3}$P_{s \rightarrow s'}^\epsilon > 0$ for some $\epsilon > 0$ $\Rightarrow$ $0 < \lim_{\epsilon \rightarrow 0^+}\frac{P_{s \rightarrow s'}^\epsilon}{\epsilon^{R(s \rightarrow s')}} < \infty$.
	\end{enumerate}
	where $R(s \rightarrow s')$ in \ref{def: RPMC3} is some nonnegative real number and is referred to as the \textit{resistance} of the transition $s \rightarrow s'$.
	\label{def: RPMC}
\end{definition}
\begin{remark}
	We would like to remark that since irreducible, a RPMP on a finite-state space has a unique stationary distribution, namely $\pi^\epsilon$, which is uniquely determined by the balance equation $\pi = \pi P$. See, e.g., \cite{rosenthal2006first} for details.
\end{remark}

Another important concept we would like to review here is the construction of a resistance tree with some associated stochastic potential.
\begin{definition}[Resistance Tree and Stochastic Potential]
	Given the state space $S$ in Definition~\ref{def: RPMC}, we construct a complete graph (i.e. a graph with an edge between any pair of its vertices) with $|S|$ vertices, each denoted as $s_i$ for $i = 1, 2, \dots, |S|$. For any directed edge $(s_i, s_j)$, we assign on it a weight of $\rho_{ij} = R(s_i \rightarrow s_j)$ where $R(\cdot)$ is the corresponding resistance. A directed path $\CP$ with length $|\CP|$ is a sequence of joint directed edges $\CP = \{s^0 \rightarrow s^1 \rightarrow \cdots, \rightarrow s^{|\CP|} \}$ and the \textit{restistence of the path} $\CP$ is defined as $R(\CP) := \sum_{k = 1}^{|\CP|} R(s^{k - 1} \rightarrow s^k)$. Let $\CT_i$ be a tree containing all $|S|$ vertices and rooted at the vertex $s_i$, such that there exists a unique directed path $\CP_{j \rightarrow i}$ from any $s_j \neq s_i$. The \textit{resistance of $\CT_i$} is defined as 
	\begin{align*}
		R(\CT_i) := \sum_{\{s \rightarrow s'\} \in \CT_i} R(s\rightarrow s').
	\end{align*}
	Among $\CT_i$ for $\forall i = 1, 2, \dots, |S|$, there exist $\CT_i^*$(s) having minimum resistance $\gamma_i = \min_{\CT_i} R(\CT_i)$. $\gamma_i$ is then referred to as \textit{stochastic potential} of state $s_i$.
	\label{def: resistance tree}
\end{definition}

The above terminologies will play key roles in our later proofs, via a powerful theorem lemma from \cite{young1993evolution}. 
\begin{theorem}[Characterization for Stochastically Stable States]
	\label{thm: characterization of stochastic-stable}
	Consider an RPMP $(Z_t^\epsilon)_{t \in \N}$ with a transition matrix $P^\epsilon$ defined on the state space $S$. The stochastically stable states are exactly $\{s_i \in S: \gamma_i = \min_{j \in \{1, 2, \dots, |S|\}} \gamma_j\}$.
\end{theorem}

\begin{proofof}{Theorem~\ref{thm: characterization of stochastic-stable}}
See \cite{young1993evolution} Lemma 1. 
\end{proofof}

\subsection{The MH-Policy in Asynchronous G-VCGs}
\label{subsec: MH-policy}
We start from the easiest setting where agents become active and update their colors one by one. A Markov chain $(Y_t)_{t \in \N}$ is established on the coloring space $C$ with each coloring as a state. The elements of the transition matrix correspond to the probability of a direct transition between the two states. Our aim is to adjust the transition probabilities of the chain (i.e. the acceptance probability of an active agent) so that stochastically stable states only contain $c^*_{WO2}$'s.

We would like to explain on some possible concerns regarding the policy before stepping towards the details in the MH-policy. In the policy, one may question that \ref{info: member} seems useless for individual decision-making. Indeed, without a loss of generality, we would restrict our consideration only on \ref{info: neighbor} (i.e. assume \ref{info: member} is not inaccessible) for the sake of simplicity in the rest of this section, which equalizes the neighborhood $\CN(V_i)$ with the family $\CF_i$. However, we are to show in Section~\ref{sec: robust G-VCGs} on why \ref{info: member} is important for some special purposes of the G-VCGs and how it can be tractably integrated in our formulations. 

The proposed MH-policy is now presented. In order to maximize the objective value in \ref{prob: optimalcoloring2}, as in the Metropolis-Hasting algorithm, we wish the acceptance probability for each active agent given the selected color $c_{new}$ to be
\begin{align}
	\alpha[(c_i(t), c_{-i}(t)) \rightarrow (c_{new}, c_{-i}(t))] = \min\{1, e^{\frac{1}{\tau}\Phi(c_{new}, c_{-i}(t)) - \Phi(c_i(t), c_{-i}(t))}\}
	\label{eqn: acceptance_ratio}
\end{align}
to satisfy the local blanced equation for the stationary distribution $\pi$, where
\begin{align*}
	\pi (c) = \frac{e^{\frac{1}{\tau} \Phi(c)}}{\sum_{\tilde{c} \in C} e^{\frac{1}{\tau} \Phi(\tilde{c})}}
\end{align*}
is a Boltzmann distribution \citep{mcquarrie2000statistical} and $\tau$ is some temperature parameter. Here we omit the terms of the proposal probability because all proposal probabilities equal in a randomized selection. Notice that, as one decreases the temperature $\tau$, the target distribution only puts weights on the state(s) where the social welfare $\Phi(\cdot)$ is maximized.

\begin{algorithm}[h]
	\KwIn{$G(V, E)$, $M$, $c(0)$, $T$, $\tau_0$}
	\KwOut{$c^*_{WO2}$ (equivalently, $c^*_{WO1}$)}
	\nl Initialize $c \leftarrow c(0)$, $t \leftarrow 0$, $\tau(0) \leftarrow \tau_0$.\\
	\nl \While{$t < T$}{
		\nl Activate $V_i \in V$ randomly. \\
		\nl Sample a color $c_{new}$ to be $c_i(t + 1)$ from the available color set $\CC_i = M$ randomly. \\
		\nl $c_{-i}(t + 1) = c_{-i}(t)$. \\
		\nl Calculate the acceptance ratio $\alpha(t)= \min\{1, e^{ \frac{1}{\tau(t)}\sum_{j \in \CF_i \cup \{V_i\}} w_j \Delta u_j ((c_i(t), c_{-i}(t)) \rightarrow (c_{new}, c_{-i}(t))} \}$.\\
		\nl Generate a number $K \in [0, 1]$ randomly.\\ 
		\nl \If{$K < \alpha(t)$}{
			\nl $c_i(t + 1) = c_{new}$. \\
			\Else{
				\nl $c_i(t + 1) = c_i(t)$.\\
			}
		}
		\nl $t\leftarrow t+1$. \\
		\nl Reduce $\tau(t - 1)$ to $\tau(t)$ under some scheme.
	}
	\nl $c^*_{WO2} \leftarrow c(T) = (c_1(T), c_2(T), \dots, c_n(T))$. \\
	\nl \Return $c^*_{WO2}$. \\
	\caption{{\bf Realization of the MH-policy in asynchronous G-VCGs}}
	\label{algo: MH-policy}
\end{algorithm}

The welfare difference in the acceptance ratio \eqref{eqn: acceptance_ratio} cannot be directly reflected from individual's utility difference as a modification on an agent's choice would likely affect his neighbors' utilities as well. Yet, we notice that, a clever way for an active agent $V_i$ to keep aware of the total difference in the entire network is to monitor the imminent local utility difference across his family $\CF_i$ before he accepts or rejects. This is because an agent choice can affect the utility of nobody other than his neighbors $V_j \in \CN(V_i)$, and we have unified $\CF_i$ and $\CN(V_i)$ to be equivalent. Formally, when $V_i$ is the unique active agent in the game, it holds that
\begin{align}
	\label{eqn: welfare-family}
	\Delta \Phi(c^k \rightarrow c^l) = \Delta U_{\CF_i \cup \{V_i\}} (c^k \rightarrow c^l)
\end{align}
where $c^k = (c_i(t), c_{-i}(t))$ and $c^l = (c_{new}, c_{-i}(t))$, and the term on the righth-and side is defined as 
\begin{align}
	\Delta U_{\CF_i\cup \{V_i\}} (c^k \rightarrow c^l) :=  \sum_{j \in \CF_i \cup \{V_i\}} w_j \Delta u_j (c^k \rightarrow c^l).
	\label{eqn: family_utility}
\end{align}
Using the terminology in a foundational work \cite{monderer1996potential}, there exists a \textit{potential} relationship between $\Phi(\cdot)$ and $U_{\CF_i\cup \{V_i\}}(\cdot)$. The procedure of realizing the MH-policy is summarized in the pseudo-code Algorithm~\ref{algo: MH-policy} in the asynchronous version.

Before moving on to prove the optimality of Algorithm~\ref{algo: MH-policy}, we would like to demonstrate why the MH-policy can be appealing to agents; in other words, it respects the human nature of \ref{feature: greedy} and \ref{feature: myopic}. This is because when updating under the MH-policy:
\begin{enumerate}[label=(\roman*)]
	\item An active agent $V_i$ will accept $c_{new}$ whenever his own utility $u_i(c_{new}, c_{-i}(t))$ increases: if $u_i(c_i(t), c_{-i}(t)) = 0$ and $u_i(c_{new}, c_{-i}(t)) > 0$ then $u_i(c_{new}, c_{-i}(t)) - u_i(c_i(t), c_{-i}(t)) < \Delta U_{\CF_i \cup \{V_i\}} (c^k \rightarrow c^l)$ since $V_i$'s zero-utility neighbors may also have the clash resolved; else, if $u_i(c_i(t), c_{-i}(t)) > 0$ and $u_i(c_{new}, c_{-i}(t)) > u_i(c_i(t), c_{-i}(t))$, then $u_i(c_{new}, c_{-i}(t)) - u_i(c_i(t), c_{-i}(t)) = \Delta U_{\CF_i \cup \{V_i\}} (c^k \rightarrow c^l)$ as the neighbors' utilities are unaffected. In both cases, $e^{\Delta U_{\CF_i \cup \{V_i\}} (c^k \rightarrow c^l)} > 1$. 
	\item \label{appealing-2} An active agent $V_i$ is not obliged to give up on his old color whenever his own utility $u_i(c_{new}, c_{-i}(t))$ decreases: if $u_i(c_i(t), c_{-i}(t)) > 0$ and $u_i(c_{new}, c_{-i}(t)) = 0$ then $u_i(c_{new}, c_{-i}(t)) - u_i(c_i(t), c_{-i}(t)) > \Delta U_{\CF_i \cup \{V_i\}} (c^k \rightarrow c^l)$ since $c_{new}$ induces color clashes between $V_i$ and some neighbors whose utility would immediately drop to zero; else, if $u_i(c_i(t), c_{-i}(t)) > 0$ and $u_i(c_{new}, c_{-i}(t)) > 0$, then $u_i(c_{new}, c_{-i}(t)) - u_i(c_i(t), c_{-i}(t)) = \Delta U_{\CF_i \cup \{V_i\}} (c^k \rightarrow c^l)$ as the neighbors' utilities are unaffected. In both cases, $e^{\Delta U_{\CF_i \cup \{V_i\}} (c^k \rightarrow c^l)} < 1$. 
	\item The more $V_i$'s utility would drop, the more likely he would keep his old color. The arguments are the same as \ref{appealing-2} along with the fact that the worst utility is zero.
\end{enumerate}

We now step to prove that the MH-policy indeed leads to $c_{WO2}^*$. To do so we first state that the Markov Chain induced by the MH-policy is in fact an RPMP.
\begin{lemma}[MH-Policy Induces RPMP]
	Let Assumptions~\ref{assp: utilitarian welfare}, \ref{assp: colors} and \ref{assp: random} hold. Consider an asynchronous G-VCG $\CG$. The MH-policy induces an RPMP where the unperturbed process only accesses to the colorings bringing in better welfares. The corresponding resistance of a state transition $c \rightarrow c'$ is 
	\begin{align*}
		R(c \rightarrow c') = \max \{0, - \Delta U_{\CF_i\cup \{V_i\}} (c \rightarrow c‘)\}
	\end{align*}
	for $c := (c_i, c_{-i})$ and $c' := (c_i', c_{-i})$ only differing on vertex $V_i$.
	\label{lemma: MH-induced RPMP}
\end{lemma}
\begin{proofof}{Lemma~\ref{lemma: MH-induced RPMP}}
See Appendix~\ref{sapx: MH-policy}. 
\end{proofof}

We are now able to invoke properties of RPMP to investigate on the support of the stationary distribution $\pi(\cdot)$, i.e. the set of stochastically stable states, which only contains states leading to the global optimal welfare rather than a local optimum. Besides, it turns out that we can establish relationship between the resistances and the final welfares. Equipped with these tools, we substantiate that the MH-policy indeed leads the game to $c^*_{WO2}$.

\begin{theorem}[The MH-Policy is Optimal]
	Let Assumptions~\ref{assp: utilitarian welfare}, \ref{assp: colors} and \ref{assp: random} hold. Consider an asynchronous G-VCG $\CG$. Then $\CG$ under the MH-policy converges to $c^*_{WO2}$, the global optimal solution to \ref{prob: optimalcoloring2} (or, equivalently, to \ref{prob: optimalcoloring1} by \eqref{optimum coincide}).
	\label{thm: MH-policy optimal}
\end{theorem}
\begin{proofof}{Theorem~\ref{thm: MH-policy optimal}}
See Appendix~\ref{sapx: MH-policy}. 
\end{proofof}

\subsection{The MH-Policy in Independently Synchronous G-VCGs}
\label{subsec: MH indep_synchronous}
We turn to the more general synchronous setting where agents may become active simultaneously. As already mentioned in Section~\ref{sec: IP-VCG}, an arbitrary synchronous G-VCG is extremely difficult to trace and optimality may be no longer guaranteed. However, inspired by \cite{marden2012revisiting}, we show in this section that when restricted to a collection of independently synchronic G-VCGs, the optimality of the MH-policy is still preserved when the probability of each agent being active is small enough.

Analogue to our discussions in Section~\ref{subsec: MH-policy}, the following Lemma~\ref{lemma: MH-induced RPMP_indep-synchronous} indicates that the MH-policy in independent synchronous settings again induces some RPMP with different resistances of state transitions. 

\begin{lemma}[MH-Policy Induces RPMP with independent synchrony]
	Let Assumptions~\ref{assp: utilitarian welfare}, \ref{assp: colors} and \ref{assp: random} hold. Consider an independently synchronous G-VCG $\CG$ with an activation parameter $\omega_t$. Define $\epsilon_t := e^{- \frac{1}{\tau(t)}}$. Then the MH-policy induces an RPMP and the unperturbed process only accesses to the colorings bringing in better welfares. The corresponding resistance of a state transition $c \rightarrow c'$ is 
	\begin{align*}
		R(c \rightarrow c') = \sum_{V_i \in G} \max \{0, -\Delta U_{\CF_i\cup \{V_i\}} ((c_i, c_{-i}) \rightarrow (c_i', c_{-i}))\}
	\end{align*}
	where $G := \{V_j: c_j \neq c_j'\}$.
	\label{lemma: MH-induced RPMP_indep-synchronous}
\end{lemma}
\begin{proofof}{Lemma~\ref{lemma: MH-induced RPMP_indep-synchronous}}
See Appendix~\ref{sapx: MH indep_synchronous}. 
\end{proofof}

It remains to check whether a stochastically stable state corresponds to a maximum welfare value. Unfortunately, this is not always the case as a G-VCG would be likely to get stuck in some local maximum state. Nevertheless, in the next Theorem~\ref{thm: optimality MH synchronous}, we will show the optimality of the MH-policy is preserved when the activation probability $\omega_t$ is small enough when $\log_{\epsilon_t} \omega_t = \CK > 0$ is constant.

\begin{theorem}[Optimality of the MH-Policy with low-$\omega$ independent synchrony]
	Let Assumptions~\ref{assp: utilitarian welfare}, \ref{assp: colors} and \ref{assp: random} hold. Consider an independently synchronic G-VCG $\CG$ with activation probability $\omega_t$. Define $\epsilon_t := e^{- \frac{1}{\tau(t)}}$. If $\log_{\epsilon_t} \omega_t = \CK$ is constant and $\CK \geq 2 \phi^*$, then the MH-policy also induces another RPMP with resistance
	\begin{align*}
		R(c \rightarrow c') = \sum_{V_i \in G} \CK |G| + \max \{0, -\Delta U_{\CF_i\cup \{V_i\}} ((c_i, c_{-i}) \rightarrow (c_i', c_{-i}))\}
	\end{align*}
	and $\CG$ under the MH-policy converges to $c^*_{WO2}$, the global optimal solution to \ref{prob: optimalcoloring2} (or, equivalently, to \ref{prob: optimalcoloring1} by \eqref{optimum coincide}).
	\label{thm: optimality MH synchronous}
\end{theorem}

\begin{proofof}{Theorem~\ref{thm: optimality MH synchronous}}
See Appendix~\ref{sapx: MH indep_synchronous}. 
\end{proofof}

\begin{remark}
	Theorem~\ref{thm: optimality MH synchronous} serves as a theoretical gurantee for the asymptotic behavior of the RPMP induced by MH policy, yet would be impractical to observe as $\omega_t$ becomes extremely small when $t \rightarrow \infty$. However, empirical evidence from experiments presented in Section~\ref{sec: simulation} show that the MH policy still performs well for independently synchronous cases with some constant $\omega$. Any relaxation on the assumption made in Theorem~\ref{thm: optimality MH synchronous} would be left to future work. 
\end{remark}

\section{Robust Welfare Optimization for G-VCGs}
\label{sec: robust G-VCGs}

In previous sections, we exhaust the information of \ref{info: neighbor} for agents' decision-making while \ref{info: member} is not fully employed. In this section, we would convince the readers on the importance of the latter in real-life events planning and how it can help with some welfare optimization in G-VCGs for additional purposes, namely \textit{robust welfare optimization} (RWO). Besides the factors introduced above, another key consideration in venue selection of events planning is the time between consecutive events conducted in the same venue. If two events arranged in the same venue have their schedules close to each other, the second one would take higher risk of being affected by a possible overrun of the previous one. Therefore, an ideal welfare function should impose penalties for colorings where two inadjacent vertices highly likely to be linked are  share the same colors. The probability of the presence of an complementary edge between vertices $V_i$ and $V_j$ can be estimated by the below formula proposed by \cite{lim2005robust}
\begin{align}
	p_{ij} \equiv \frac{\tilde{t}}{\max\{T_i^s, T_j^s\} - \min\{T_i^e, T_j^e\} + \alpha}.
\end{align}
where $\tilde{t}$ is the minimum event transfer time, $(T^s_i, T^e_i)$ and $(T^s_j, T^e_j)$ are the starting and ending time for the two events, and $\alpha$ is some predefined constant. In the rest of this section, we assume such probabilities are public information in each family.

The idea of RWO comes from the concept of robust coloring problem (RCP) firstly proposed in \cite{yanez2003robust} and soon gained numerous popularities; see, e.g., \cite*{lim2005robust, wang2013metaheuristics, archetti2014branch}. All these literatures analyzed the robust coloring as a separate optimization problem while, to our knowledge, have not considered its iteraction with other objectives. We therefore fill the gap to include in our discussion the robust welfare optimization as an ``extension'' to the baseline \ref{prob: optimalcoloring1} and \ref{prob: optimalcoloring2}, thus control the social welfare with more influencing factors in the network. By ``extension'', we will explain later that $c^*_{WO1}$ would be an input of our algorithm to solve RWO which should have been figured out in Algorithm~\ref{algo: MH-policy}; i.e. one can interpret RWO as a second-stage problem.

\subsection{Motivation and Formulation}
\label{subsec: motivation_formulation_rwo}
We would like to give a brief overview on the main idea behind the RCP before formally introducing our RWO problem. Besides requiring the coloring to be proper as classical graph coloring does, a robust coloring takes into account the complementary edges in a network. Given each complementary edge a probability to be connected, the objective is to fathom out a most ``robust'' coloring in a sense that the probability for the properness to be destroyed by connecting one or more complementary edges is minimized. The number of colors $m$ is fixed as a constraint. An RCP is formulated as follows:
\begin{align}
	\begin{split}
		\minimize_{c \in C} \quad & \sum_{\{V_i, V_j\} \in \bar{E}; c_i = c_j} p_{ij} \\
		\mathrm{subject~to} \quad & c_i \neq c_j, \forall i, \forall j \in \{j: \CL_{ij} = 1\}
	\end{split}
	\label{prob: RCP}
	\tag{RCP}
\end{align} 
where $p_{ij}$ is the probability for a complementary $\{V_i, V_j\}$ to be connected.

The literature works as mentioned explored multifarious applications which can be modelled by \ref{prob: RCP}, which simply assumed an equal ``damage'' brought by any new pair-connection, to our knowledge. Note that the formulation of \ref{prob: RCP} reflects on this assumption by giving each pair-connection a damage of 1, without a loss of generality. This is, however, not true in our setting of G-VCGs with utility variations, since the amount of welfare can be affected differently when a different complementary edge is connected. Once a complementary edge is connected between two vertices in the same color, the utilities of both get eliminated from the total social welfare as a new clash occurs. Indeed in reality, robustness is most times not the only pursuit and there is always a trade-off between welfare and risk (i.e. decreasing robustness). To integrate the consideration of such robustness into our welfare optimization problem, we add in \ref{prob: optimalcoloring1} an ``expected loss'' term $\expectation(\FL)$ to be minimized. This is inspired by the spirit of the general stochastic optimization to quantify the average uncertainty. We formulate our RWO problem to be
\begin{align}
	\begin{split}
		\maximize_{c \in C} \quad & \Phi_R(c) = \sum_{V_i \in V} w_i u_i(c) - \expectation(\FL) \\
		\mathrm{subject~to} \quad & c_i \neq c_j, \forall i, \forall j \in \{j: \CL_{ij} = 1\}.
	\end{split}
	\label{prob: RWO}
	\tag{RWO}
\end{align}

Suppose each complementary edge has a probability $p_{ij}$ to be connected. Denote $\FN \subseteq \bar{E}$ to be the set of new-connected complementary edges. Let $\FC_i := \{ \{V_i, V_j\} \in \bar{E}: V_j \in V, c_i = c_j\}$ be the set of complementary edges taking $V_i$ as an endpoint and with endpoints' colors identical . It is important to note that $\FL$ in \ref{prob: RWO} is not a simple summation of the utility loss in both endpoints from a pair-connection, otherwise the losses would be counted repeatedly. For example, after the utility of $V_i \in V$ is eliminated due to a new connected clash, another complementary edge with him as an endpoint would no longer reduce its utility. Instead, we may express $\FL$ as a weighted sum of $n$ random variables $X_1, X_2, \dots, X_n$ defined by 
\begin{align*}
	X_i = u_i(c) \indicator_{\FC_i \cap \FN \neq \emptyset}, \quad \forall i \in \{1, 2, \dots, n\}
\end{align*}
thus
\begin{align}
	\expectation(\FL) = \sum_{i = 1}^{n} w_i \expectation[u_i(c) \indicator_{\FC_i \cap \FN \neq \emptyset}] =  \sum_{i = 1}^{n} w_i u_i(c) \Pr(\FC_i \cap \FN \neq \emptyset).
	\label{eqn: expected_loss}
\end{align}

\vspace{10pt}
\subsection{The MH-Policy for RWO}
\label{subsec: MH_RWO}
In this subsection, we again employ the Metropolis-Hasting principle to solve \ref{prob: RWO} in a decentralized fashion. A key difference to the MH-policy in Section~\ref{sec: MH-policy} is that the family structure is dynamic when \ref{info: member} is involved. Note that the proposed MH-policy for \ref{prob: RWO} is continued as an extension to solving \ref{prob: optimalcoloring1}, thus taking the output of Algorithm~\ref{algo: MH-policy} as input.
\begin{assumption}
	We replace Assumption~\ref{assp: random1} by the following statement while keep \ref{assp: random2} and \ref{assp: random3} hold.
	\begin{enumerate}[label=\normalfont{(R\arabic*-RWO)},leftmargin=60pt]
		\item \label{assp: random1_rwo}An active agent, say $V_i$, would first randomly select a color $c_{new}$ from the color set $M / c_{\CN(V_i)}$; i.e. $\CC_i = M / c_{\CN(V_i)}$ in each round. 
	\end{enumerate}
	\label{assp: random_rwo}
\end{assumption}

\begin{remark}
	We would like to clarify on some possible doubts regarding our design:
	\begin{enumerate}[label=(\roman*)]
		\item Note that Assumption~\ref{assp: random_rwo} is not a loss of generality. An equivalent approach is to work with Assumption~\ref{assp: random} and set the loss for an improper coloring to be $-\infty$ so that the game never accepts improper colorings if starting from $c^*_{WO1}$ which is proper. In other words, instead of a change in the general setting, Assumption~\ref{assp: random_rwo} merely allows an acceleration compared to Assumption~\ref{assp: random}.
		\item The reason why we start our algorithm from $c^*_{WO1}$ rather than thoroughly starting from the very beginning arbitrary coloring $c(0)$ is to guarantee the properness of our optimal coloring. Note that if we use the relaxation in \ref{prob: optimalcoloring2} in solving \ref{prob: RWO}, the optimal coloring may not be proper; for example, when a vertex's utility is less than the possible loss that might be triggered. Starting from $c^*_{WO1}$ along with Assumption~\ref{assp: random_rwo} makes sure the game keeps running on proper colorings.
	\end{enumerate}
\end{remark}

\begin{algorithm}[h!]
	\KwIn{$G(V, E)$, $M$, $c^*_{WO1}$ (equivalently $c^*_{WO2}$), $T$, $\tau_0$}
	\KwOut{$c^*_{RWO}$}
	\nl Initialize $c \leftarrow c^*_{WO1}$, $t \leftarrow 0$, $\tau(0) \leftarrow \tau_0$.\\
	\nl \While{$t < T$}{
		\nl Activate $V_i \in V$ randomly. \\
		\nl Sample a color $c_{new}$ from the available color set $\CC_i = M / c_{\CN(V_i)}$ randomly. \\
		\nl $c_{-i}(t + 1) = c_{-i}(t)$. \\
		\nl Calculate the acceptance ratio $\alpha(t) = \min\{1, e^{ \frac{1}{\tau(t)}[\Delta U_{\CN(V_i) \cup \{V_i\}} (c^k \rightarrow c^l) - \Delta L_{\CF_i \cup \{V_i\}/ \CN(V_i)}  (c^k \rightarrow c^l)]} \}$.\\
		\nl Generate a number $K \in [0, 1]$ randomly.\\ 
		\nl \If{$K < \alpha(t)$}{
			\nl $c_i(t + 1) = c_{new}$. \\
			\Else{
				\nl $c_i(t + 1) = c_i(t)$.\\
			}
		}
		\nl $t\leftarrow t+1$. \\
		\nl Reduce $\tau(t - 1)$ to $\tau(t)$ under some scheme.
	}
	\nl $c^*_{RWO} \leftarrow c(T) = (c_1(T), c_2(T), \dots, c_n(T))$. \\
	\nl \Return $c^*_{RWO}$. \\
	\caption{{\bf Realization of the MH-policy for \ref{prob: RWO} in asynchronous G-VCGs}}
	\label{algo: MH-policy_RWO}
\end{algorithm}

As in Section~\ref{sec: MH-policy}, we would first consider asynchronous cases for solving \ref{prob: RWO}. Denote by $\FL_{\FS^{(c)}}$ the loss occurs in a set of vertices $\FS \subseteq V$ under a coloring $c$. When the coloring is changed from $c^k$ to $c^l$ in round $t$ where $c^k = (c_i(t), c_{-i}(t))$ and $c^l = (c_{new}, c_{-i}(t))$, the difference in the total welfare $\Phi_R$ would be
\begin{align}
	\begin{split}
		\Delta \Phi_R(c^k \rightarrow c^l) &= \Delta U_{\CF_i \cup \{V_i\}} (c^k \rightarrow c^l) - \Delta L_{\CF_i \cup \{V_i\}}  (c^k \rightarrow c^l) \\
		&= \Delta U_{\CN(V_i) \cup \{V_i\}} (c^k \rightarrow c^l) - \Delta L_{\CF_i \cup \{V_i\}/ \CN(V_i)}  (c^k \rightarrow c^l)
	\end{split}
	\label{eqn: Delta_Phi_R}
\end{align}
where $\Delta U_{\CN(V_i) \cup \{V_i\}} (c^k \rightarrow c^l)$ is defined as in \ref{eqn: family_utility} and
\begin{align}
	\Delta L_{\CF_i \cup \{V_i\}/ \CN(V_i)} (c^k \rightarrow c^l) := \expectation[\FL_{\CF_i^{(c)}\cup \{V_i\}/ \CN(V_i)} | c = c^l] - \expectation[\FL_{\CF_i^{(c)}\cup \{V_i\}/ \CN(V_i)} | c = c^k].
	\label{eqn: loss_def}
\end{align}
The second equality in \eqref{eqn: Delta_Phi_R} holds because a color change in $V_i$ will only affect the utilities of his neighbors sharing \ref{info: neighbor} and only affect the underlying loss of his non-neighboring family members sharing \ref{info: member}, and $c_i \notin c_{\CN(V_i)}$ by Assumption~\ref{assp: random_rwo}. Since the probability terms in \eqref{eqn: expected_loss} can be expressed using $p_{ij}$ by 
\begin{align}
	\begin{split}
		\Pr(\FC_i \cap \FN \neq \emptyset) &= 1 - \Pr(\FC_i \cap \FN = \emptyset) \\
		&= 1 - \prod_{j: \{V_i, V_j\} \in \bar{E}, c_i = c_j} (1 - p_{ij})
	\end{split}
	\label{eqn: prob}
\end{align}
and vertices other than $V_i$ keep their colors unchanged, we can decompose \ref{eqn: loss_def} as 
\begin{align}
	\begin{split}
		\Delta L_{\CF_i \cup \{V_i\}/ \CN(V_i)} (c^k \rightarrow c^l)&= w_i \biggl[u_i(c^l)\biggl(1 - \prod_{j: V_j \in \CF_i^{(c^l)} / \CN(V_i)} (1 - p_{ij})\biggl) - u_i(c^k)\biggl (1 - \prod_{\tilde{j}: V_{\tilde{j}} \in \CF_i^{(c^k)}/ \CN(V_i)}  (1 - p_{i \tilde{j}})\biggl) \biggl] \\
		& \quad \quad - \sum_{\tilde{j}: V_{\tilde{j}} \in  \CF_i^{(c^k)}/ \CN(V_i)} p_{i\tilde{j}} w_{\tilde{j}} u_{\tilde{j}}(c^k)\\
		& \quad \quad + \sum_{j: V_j \in \CF_i^{(c^l)} / \CN(V_i)} p_{ij}w_j u_j(c^l)
	\end{split}
	\label{eqn: decomposition_loss}.
\end{align}
Notice that the terms included in \ref{eqn: decomposition_loss} are all accessible to $V_i$ during his color updating procedure, since losses can only be triggered by vertices in the same color which are all included in the family. Therefore, \ref{eqn: Delta_Phi_R} can be calculated by each decentralized active agent. The pseudo-code for realizing the MH policy for \ref{prob: RWO} is provided in Algorithm~\ref{algo: MH-policy_RWO}. 

\begin{remark}
	Note that the probability of a connection in any complementary edge whose both endpoints share the same color is known to the corresponding vertices, since it belongs to \ref{info: member}.
\end{remark}

The optimality of the MH-policy for \ref{prob: RWO} in either asynchronous and independently synchronous settings can be again substantiated using theories of RPMP. Arguments are similar to the ones in Section~\ref{sec: MH-policy} though associated with some other resistance with a minor modification, and we will not repeat the procedure here.

\section{Simulation Experiments}
\label{sec: simulation}
We examine the effectiveness of the proposed MH-policies  for G-VCGs in this section. The probabilitsitic network stuctures were constructed using the Erdős–Rényi model provided by the \textit{networkx} package in Python with certain connection probabilities. Meanwhile, we randomly generated preferences from $(0, 100)$ for each vertex as well as weights from $(0, 1)$ which was further normalized to sum up to 1. We monitor the dynamics induced by the networks with different sizes, different connection probabilities (thus different $\delta$) and different temperature reduction schemes. We also checked on the effects of varied activation parameters $\omega$ in synchronous settings to see how synchrony affects the efficiency of the MH-policy driven algorithms. For comparison purposes, we also implemented a tabu search (TS) algorithm adapted from \cite{lim2005robust} as an aid to determine whether the obtained welfares are optimal (or, at least, sub-optimal), of which the details are given in Algorithm~\ref{algo: tabu_search} in Section~\ref{sec: tabu}.

\begin{table}[htbp]
	\centering
	\begin{tabular}{|c|c|c|}
		\hline
		\textbf{Temperature Scheme} & \textbf{Formula} & $\tau(0)$ \\
		\hline
		constant & $\tau(t) = \tau(0)$ & 0.01\\
		\hline
		exponential multiplicative & $\tau(t) = 0.99^t \tau(0)$  & 10\\ 
		\hline
		logarithmical & $\tau(t) = \tau(0) / 1 + \log(100 + t)$ & 0.1\\
		\hline
		trigonometric additive & $\tau(0) = 0.01 + 0.05(\tau(0) - 0.01)(1 + \cos(\frac{t \pi}{T})) $& 10\\
		\hline
	\end{tabular}
	\caption{Different temperature schemes}
	\label{table: temperature}
\end{table}

Four temperature reduction schemes were considered in our experiments: constant, exponential multiplicative, logarithmical and trigonometric additive. The specific parameters are given in Table~\ref{table: temperature}. The experiements were divided into several groups, each of which has one monitored variable with other variables controlled. For each group, we ran three parallel simulations and would present one of them in this section. The number of iterations in each simulation experiment was set to be $T = 1 \times 10^4$. 

\begin{figure}[h!]
	\centering
	\begin{minipage}{.48\textwidth}
		\centering
		\includegraphics[width = \linewidth]{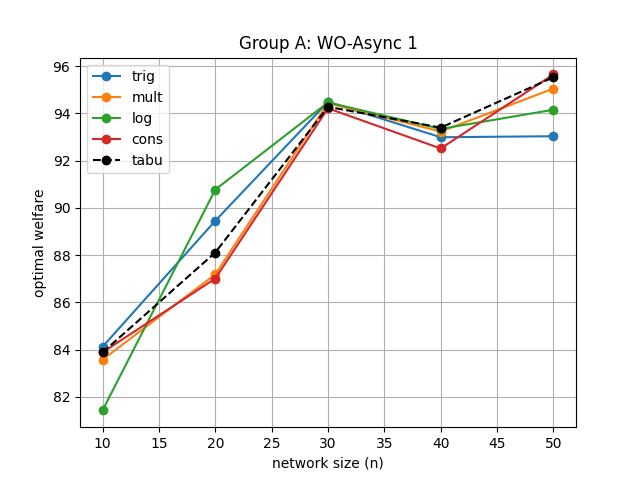}
		\caption{Optimal Welfare Comparison for Group A}
		\label{fig: groupA}
	\end{minipage}
	\hspace{2mm}
	\begin{minipage}{.48\textwidth}
		\centering
		\includegraphics[width = \linewidth]{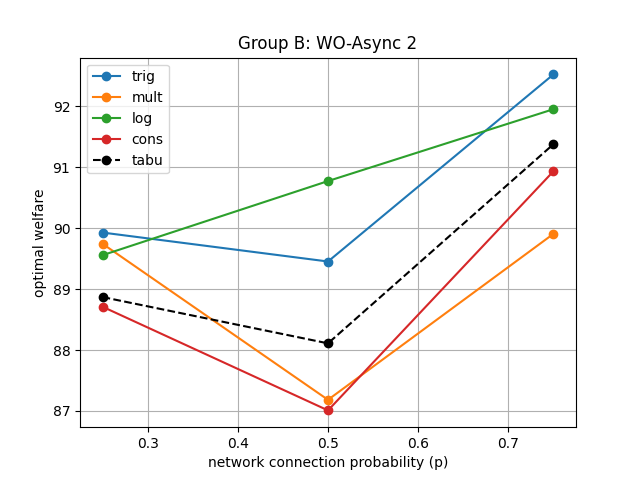}
		\caption{Optimal Welfare Comparison for Group B}
		\label{fig: groupB}
	\end{minipage}
	\vspace{2mm}
	\begin{minipage}{.48\textwidth}
		\centering
		\includegraphics[width = \linewidth]{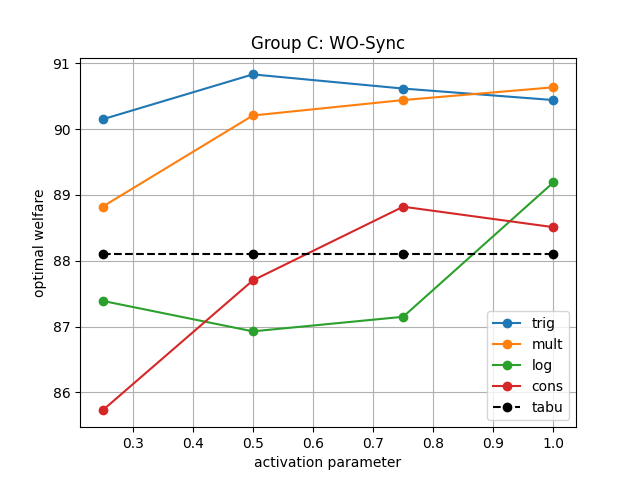}
		\caption{Optimal Welfare Comparison for Group C}
		\label{fig: groupC}
	\end{minipage}
\end{figure}

Table~\ref{table: WO} gives the results for the three groups which focus on solving \ref{prob: optimalcoloring2}. Among the data given by the four temperature schemes, the welfares outperforming TS are labelled in red while the ones in blue suggest that the difference between it and the TS outcome is less than 0.5. The experiments in the first two groups were all conducted in an asynchronous manner as described in Section~\ref{subsec: setting}. The group ``WO-Async 1'' aims to detect on the influence made by variations of the network size. It can be observed from Figure~\ref{fig: groupA} that the MH-policy supported by the logarimical and trigonometric addtive schemes achieves higher welfares on this task than TS did, and all methods perform equally well in general. In terms of group ``WO-Async 2'', we are able to analyze on the impacts under different network degrees by adjusting the priori connection probability between edges when generating the graph. It turned out that trigonometric additive and logarithmical models still prevail over TS as can be observed from Figure~\ref{fig: groupB}. 
\begin{table}[h!]
	\centering
	\begin{tabular}{||c|c|cccc|c||}
		\hline
		\textbf{Group} & \textbf{Network Features} & \textbf{Trig} & \textbf{Exp} & \textbf{Log} & \textbf{Const} & \textbf{TS} \\
		\hline
		\multirow{5}{*}{WO-Async 1} & $n = 10$, $p = 0.5$ & \SC{84.12} & \CZY{83.57} & 81.43 & \CZY{83.88} &  \textbf{83.88} \\
		\cline{2-7}
		& $n = 20$, $p = 0.5$ &\SC{89.45} & 87.18 & \SC{90.77} & 87.01 &  \textbf{88.11} \\
		\cline{2-7}
		& $n = 30$, $p = 0.5$ & \SC{94.49} & \SC{94.44} & \SC{94.47} & \CZY{94.21} & \textbf{94.28} \\
		\cline{2-7}
		& $n = 40$, $p = 0.5$ &\CZY{92.99} & \CZY{93.23} & \CZY{93.34} & \CZY{92.52} &  \textbf{93.40} \\
		\cline{2-7}
		& $n = 50$, $p = 0.5$ &93.03 & \CZY{95.05} & 94.16 & \SC{95.65} &  \textbf{95.53} \\
		\hline \hline
		\multirow{3}{*}{WO-Async 2} & $n = 20, p = 0.25$ & \SC{89.93} & \SC{89.74} & \SC{89.56} & 88.71 & \textbf{88.87} \\
		\cline{2-7}
		& $n = 20, p = 0.5$  & \SC{89.45} & 87.18 & \SC{90.77} & 87.01 & \textbf{88.11} \\
		\cline{2-7}
		& $n = 20, p = 0.75$& \SC{92.53} & 89.91 & \SC{91.95} & \CZY{90.94} & \textbf{91.38} \\
		\hline \hline
		\multirow{4}{*}{WO-Sync } & $n = 20, p = 0.5, \omega = 0.25$ & \SC{90.15} & \SC{88.82} & 87.39 & 85.73 & \textbf{88.11} \\
		\cline{2-7}
		& $n = 20, p = 0.5, \omega = 0.5$  & \SC{90.83} & \SC{90.21} & 86.93 & \CZY{87.70} & \textbf{88.11} \\
		\cline{2-7}
		& $n = 20, p = 0.5, \omega = 0.75$& \SC{90.62} & \SC{90.44} & 87.15 & \SC{88.82} & \textbf{88.11} \\
		\cline{2-7}
		& $n = 20, p = 0.5, \omega = 1$& \SC{90.44} & \SC{90.63} & \SC{89.19} & \SC{88.51} & \textbf{88.11} \\
		\hline
	\end{tabular}
	\caption{Simulation results for \ref{prob: optimalcoloring2}-solving.}
	\label{table: WO}
\end{table}

The MH-policies under synchronous settings were examined in the third group, namely ``WO-Sync''. It is uplifting to witness that the proposed MH-policies can almost achieve more optimal solutions than TS, especially when driven by trignometric addtive and exponential multiplicative schemes. It is also remarkable that the algorithms are fairly robust even under completely synchronous cases. A visualization for this is given in Figure~\ref{fig: groupC}. 

\begin{figure}[h!]
	\centering
	\begin{minipage}{.48\textwidth}
		\centering
		\includegraphics[width = \linewidth]{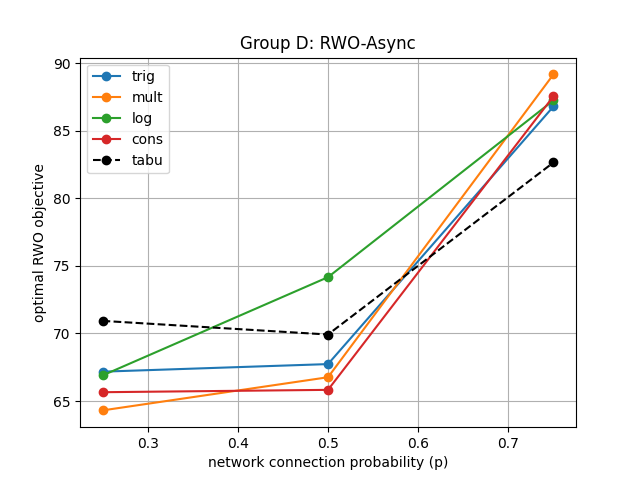}
		\caption{Optimal Welfare Comparison for Group D}
		\label{fig: groupD}
	\end{minipage}
	\hspace{2mm}
	\begin{minipage}{.48\textwidth}
		\centering
		\includegraphics[width = \linewidth]{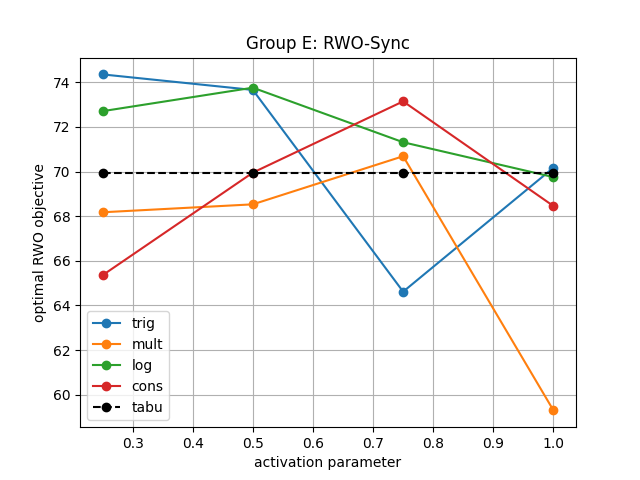}
		\caption{Optimal Welfare Comparison for Group E}
		\label{fig: groupE}
	\end{minipage}
\end{figure}

We further proceed our investigation to solving \ref{prob: RWO}, the results of which are reported in Table~\ref{table: RWO}. Due to our formulation of the expected loss term in \ref{eqn: expected_loss}, the final objective highly depends on the probability of a future connection of each complementary edge, which is directly related to the degree of the network. Therfore, in the fourth ``RWO-Async'' group, we had the network size controlled and modified the connection probability $p$, as did in ``WO-Async 2''. It turned out that the MH-policy driven Algorithm~\ref{algo: MH-policy_RWO} performs well when $p$ is relatively large, no matter what scheme it employed, as shown in Figure~\ref{fig: groupD}A possible explanation for this phenomenon is that, when $p$ is high, the number of complementary edges falls thus the optimal solution of \ref{prob: RWO} stays close $c_{WO2}^*$, thus the two-stage MH-policy becomes advantageous. Figure~\ref{fig: groupE} corresponds to the last group which again focused on the impact of synchrony. Like in ``WO-Sync'', the effectiveness of the proposed MH-policy confronted by different activation parameters are also well demonstrated. 

\begin{table}[htbp]
	\centering
	\begin{tabular}{||c|c|cccc|c||}
		\hline
		\textbf{Group} & \textbf{Network Features} & \textbf{Trig} & \textbf{Exp} & \textbf{Log} & \textbf{Const} & \textbf{TS} \\
		\hline
		\multirow{3}{*}{RWO-Async} & $n = 20, p = 0.25$ & 67.17 & 64.30 & 66.89 & 65.65 & \textbf{70.93} \\
		\cline{2-7}
		& $n = 20, p = 0.5$  & 67.73& 66.76 & \SC{74.16} & 65.83 & \textbf{69.92} \\
		\cline{2-7}
		& $n = 20, p = 0.75$& \SC{86.80} & \SC{89.17} & \SC{87.25} & \SC{87.55} & \textbf{82.65} \\
		\hline \hline
		\multirow{4}{*}{RWO-Sync } & $n = 20, p = 0.5, \omega = 0.25$& \SC{74.35} & 68.17 & \SC{72.70} & 65.35 & \textbf{69.92} \\
		\cline{2-7}
		& $n = 20, p = 0.5, \omega = 0.5$& \SC{73.66} & 68.53 & \SC{73.76}& \SC{69.95} & \textbf{69.92} \\
		\cline{2-7}
		& $n = 20, p = 0.5, \omega = 0.75$& 64.62 & \SC{70.69} & \SC{71.31} & \SC{73.14} & \textbf{69.92} \\
		\cline{2-7}
		& $n = 20, p = 0.5, \omega = 1$& \SC{70.17} & 59.30 & \CZY{69.75} & 68.45 & \textbf{69.92} \\
		\hline
	\end{tabular}
	\caption{Simulation results for \ref{prob: RWO}-solving.}
	\label{table: RWO}
\end{table}

\section{Conclusion}
\label{sec: conclusion}
In this paper, we focus on the venue selection problem commonly confronted in event planning, and model its dynamics by the proposed generalized vertex coloring games which go beyond the basic requirement of properness in search of some max-welfare coloring. It was shown that in an IP-VCG, being completely greedy and completely synchronous does not prevent agents reaching the optimal coloring asymptotically, and the converging time was shown to be stochastically bounded by $\CO_p(\log n)$. In more general G-VCGs in asynchronous manner, there is still some policy driven by the Metropolis-Hasting algorithm which enables the self-organized agents to reach the optimal coloring without violating their greediness. The optimality of the MH policy even holds in some particular independently synchronous settings. Finally, we integrate the idea of robust coloring into our formulation seeking for a balance between the welfare and the risk, and a corresponding adaptive MH-policy for the robust welfare optimization problem is provided. The effectiveness of the proposed policies are substantiated in our simulation experiments. 

\section{Acknowledgement}
\label{sec: acknowledgement}
This project was supported by Nanyang Technological University under the URECA Undergraduate Research Programme.

\bibliographystyle{elsarticle-harv} 
\bibliography{references}

\begin{thebibliography}{47}
\expandafter\ifx\csname natexlab\endcsname\relax\def\natexlab#1{#1}\fi
\providecommand{\url}[1]{\texttt{#1}}
\providecommand{\href}[2]{#2}
\providecommand{\path}[1]{#1}
\providecommand{\DOIprefix}{doi:}
\providecommand{\ArXivprefix}{arXiv:}
\providecommand{\URLprefix}{URL: }
\providecommand{\Pubmedprefix}{pmid:}
\providecommand{\doi}[1]{\href{http://dx.doi.org/#1}{\path{#1}}}
\providecommand{\Pubmed}[1]{\href{pmid:#1}{\path{#1}}}
\providecommand{\bibinfo}[2]{#2}
\ifx\xfnm\relax \def\xfnm[#1]{\unskip,\space#1}\fi
\bibitem[{Ahmed(2012)}]{ahmed2012applications}
\bibinfo{author}{Ahmed, S.}, \bibinfo{year}{2012}.
\newblock \bibinfo{title}{Applications of graph coloring in modern computer
  science}.
\newblock \bibinfo{journal}{International Journal of Computer and Information
  Technology} \bibinfo{volume}{3}, \bibinfo{pages}{1--7}.
\bibitem[{Allen et~al.(2022)Allen, Harris, Jago, Tantrai, Jonson and
  D'Arcy}]{allen2022festival}
\bibinfo{author}{Allen, J.}, \bibinfo{author}{Harris, R.},
  \bibinfo{author}{Jago, L.}, \bibinfo{author}{Tantrai, A.},
  \bibinfo{author}{Jonson, P.}, \bibinfo{author}{D'Arcy, E.},
  \bibinfo{year}{2022}.
\newblock \bibinfo{title}{Festival and special event management}.
\newblock \bibinfo{publisher}{John Wiley \& Sons}.
\bibitem[{Archetti et~al.(2014)Archetti, Bianchessi and
  Hertz}]{archetti2014branch}
\bibinfo{author}{Archetti, C.}, \bibinfo{author}{Bianchessi, N.},
  \bibinfo{author}{Hertz, A.}, \bibinfo{year}{2014}.
\newblock \bibinfo{title}{A branch-and-price algorithm for the robust graph
  coloring problem}.
\newblock \bibinfo{journal}{Discrete Applied Mathematics}
  \bibinfo{volume}{165}, \bibinfo{pages}{49--59}.
\bibitem[{Bermond et~al.(2019)Bermond, Chaintreau, Ducoffe and
  Mazauric}]{bermond2019long}
\bibinfo{author}{Bermond, J.C.}, \bibinfo{author}{Chaintreau, A.},
  \bibinfo{author}{Ducoffe, G.}, \bibinfo{author}{Mazauric, D.},
  \bibinfo{year}{2019}.
\newblock \bibinfo{title}{How long does it take for all users in a social
  network to choose their communities?}
\newblock \bibinfo{journal}{Discrete Applied Mathematics}
  \bibinfo{volume}{270}, \bibinfo{pages}{37--57}.
\bibitem[{Carosi et~al.(2019)Carosi, Fioravanti, Gual{\`a} and
  Monaco}]{carosi2019coalition}
\bibinfo{author}{Carosi, R.}, \bibinfo{author}{Fioravanti, S.},
  \bibinfo{author}{Gual{\`a}, L.}, \bibinfo{author}{Monaco, G.},
  \bibinfo{year}{2019}.
\newblock \bibinfo{title}{Coalition resilient outcomes in max k-cut games}, in:
  \bibinfo{booktitle}{SOFSEM 2019: Theory and Practice of Computer Science:
  45th International Conference on Current Trends in Theory and Practice of
  Computer Science, Nov{\`y} Smokovec, Slovakia, January 27-30, 2019,
  Proceedings}, \bibinfo{organization}{Springer}. pp. \bibinfo{pages}{94--107}.
\bibitem[{Chaudhuri et~al.(2008)Chaudhuri, Chung~Graham and
  Jamall}]{chaudhuri2008network}
\bibinfo{author}{Chaudhuri, K.}, \bibinfo{author}{Chung~Graham, F.},
  \bibinfo{author}{Jamall, M.S.}, \bibinfo{year}{2008}.
\newblock \bibinfo{title}{A network coloring game}, in:
  \bibinfo{booktitle}{Internet and Network Economics: 4th International
  Workshop, WINE 2008, Shanghai, China, December 17-20, 2008. Proceedings 4},
  \bibinfo{organization}{Springer}. pp. \bibinfo{pages}{522--530}.
\bibitem[{Chen et~al.(2019)Chen, Delcourt, Moitra, Perarnau and
  Postle}]{chen2019improved}
\bibinfo{author}{Chen, S.}, \bibinfo{author}{Delcourt, M.},
  \bibinfo{author}{Moitra, A.}, \bibinfo{author}{Perarnau, G.},
  \bibinfo{author}{Postle, L.}, \bibinfo{year}{2019}.
\newblock \bibinfo{title}{Improved bounds for randomly sampling colorings via
  linear programming}, in: \bibinfo{booktitle}{Proceedings of the Thirtieth
  Annual ACM-SIAM Symposium on Discrete Algorithms},
  \bibinfo{organization}{SIAM}. pp. \bibinfo{pages}{2216--2234}.
\bibitem[{Cui et~al.(2008)Cui, Qin, Liu, Wang, Zhang and Cao}]{cui2008modified}
\bibinfo{author}{Cui, G.}, \bibinfo{author}{Qin, L.}, \bibinfo{author}{Liu,
  S.}, \bibinfo{author}{Wang, Y.}, \bibinfo{author}{Zhang, X.},
  \bibinfo{author}{Cao, X.}, \bibinfo{year}{2008}.
\newblock \bibinfo{title}{Modified pso algorithm for solving planar graph
  coloring problem}.
\newblock \bibinfo{journal}{Progress in Natural Science} \bibinfo{volume}{18},
  \bibinfo{pages}{353--357}.
\bibitem[{Dowsland and Thompson(2008)}]{dowsland2008improved}
\bibinfo{author}{Dowsland, K.A.}, \bibinfo{author}{Thompson, J.M.},
  \bibinfo{year}{2008}.
\newblock \bibinfo{title}{An improved ant colony optimisation heuristic for
  graph colouring}.
\newblock \bibinfo{journal}{Discrete Applied Mathematics}
  \bibinfo{volume}{156}, \bibinfo{pages}{313--324}.
\bibitem[{Enemark et~al.(2011)Enemark, McCubbins, Paturi and
  Weller}]{enemark2011does}
\bibinfo{author}{Enemark, D.P.}, \bibinfo{author}{McCubbins, M.D.},
  \bibinfo{author}{Paturi, R.}, \bibinfo{author}{Weller, N.},
  \bibinfo{year}{2011}.
\newblock \bibinfo{title}{Does more connectivity help groups to solve social
  problems}, in: \bibinfo{booktitle}{Proceedings of the 12th ACM conference on
  Electronic commerce}, pp. \bibinfo{pages}{21--26}.
\bibitem[{Fleurent and Ferland(1996)}]{fleurent1996genetic}
\bibinfo{author}{Fleurent, C.}, \bibinfo{author}{Ferland, J.A.},
  \bibinfo{year}{1996}.
\newblock \bibinfo{title}{Genetic and hybrid algorithms for graph coloring.}
\newblock \bibinfo{journal}{Annals of operations research}
  \bibinfo{volume}{63}.
\bibitem[{Fryganiotis et~al.(2023)Fryganiotis, Papavassiliou and
  Pelekis}]{fryganiotis2023note}
\bibinfo{author}{Fryganiotis, N.}, \bibinfo{author}{Papavassiliou, S.},
  \bibinfo{author}{Pelekis, C.}, \bibinfo{year}{2023}.
\newblock \bibinfo{title}{A note on the network coloring game: A randomized
  distributed ($\delta$+ 1)-coloring algorithm}.
\newblock \bibinfo{journal}{Information Processing Letters}
  \bibinfo{volume}{182}, \bibinfo{pages}{106385}.
\bibitem[{Goonewardena et~al.(2014)Goonewardena, Akbari, Ajib and
  Elbiaze}]{goonewardena2014minimum}
\bibinfo{author}{Goonewardena, M.}, \bibinfo{author}{Akbari, H.},
  \bibinfo{author}{Ajib, W.}, \bibinfo{author}{Elbiaze, H.},
  \bibinfo{year}{2014}.
\newblock \bibinfo{title}{On minimum-collisions assignment in heterogeneous
  self-organizing networks}, in: \bibinfo{booktitle}{2014 IEEE Global
  Communications Conference}, \bibinfo{organization}{IEEE}. pp.
  \bibinfo{pages}{4665--4670}.
\bibitem[{Hamza et~al.(2021)Hamza, Toonsi and Shamma}]{hamza2021metropolis}
\bibinfo{author}{Hamza, D.}, \bibinfo{author}{Toonsi, S.},
  \bibinfo{author}{Shamma, J.S.}, \bibinfo{year}{2021}.
\newblock \bibinfo{title}{A metropolis-hastings algorithm for task allocation},
  in: \bibinfo{booktitle}{2021 60th IEEE Conference on Decision and Control
  (CDC)}, \bibinfo{organization}{IEEE}. pp. \bibinfo{pages}{4539--4545}.
\bibitem[{Hastings(1970)}]{hastings1970monte}
\bibinfo{author}{Hastings, W.K.}, \bibinfo{year}{1970}.
\newblock \bibinfo{title}{Monte carlo sampling methods using markov chains and
  their applications} .
\bibitem[{Hayes et~al.(2007)Hayes, Vera and Vigoda}]{hayes2007randomly}
\bibinfo{author}{Hayes, T.P.}, \bibinfo{author}{Vera, J.C.},
  \bibinfo{author}{Vigoda, E.}, \bibinfo{year}{2007}.
\newblock \bibinfo{title}{Randomly coloring planar graphs with fewer colors
  than the maximum degree}, in: \bibinfo{booktitle}{Proceedings of the
  thirty-ninth annual ACM symposium on Theory of computing}, pp.
  \bibinfo{pages}{450--458}.
\bibitem[{Hayes and Vigoda(2006)}]{hayes2006coupling}
\bibinfo{author}{Hayes, T.P.}, \bibinfo{author}{Vigoda, E.},
  \bibinfo{year}{2006}.
\newblock \bibinfo{title}{Coupling with the stationary distribution and
  improved sampling for colorings and independent sets} .
\bibitem[{Hern{\'a}ndez and Blum(2012)}]{hernandez2012distributed}
\bibinfo{author}{Hern{\'a}ndez, H.}, \bibinfo{author}{Blum, C.},
  \bibinfo{year}{2012}.
\newblock \bibinfo{title}{Distributed graph coloring: an approach based on the
  calling behavior of japanese tree frogs}.
\newblock \bibinfo{journal}{Swarm Intelligence} \bibinfo{volume}{6},
  \bibinfo{pages}{117--150}.
\bibitem[{Hindi and Yampolskiy(2012)}]{hindi2012genetic}
\bibinfo{author}{Hindi, M.M.}, \bibinfo{author}{Yampolskiy, R.V.},
  \bibinfo{year}{2012}.
\newblock \bibinfo{title}{Genetic algorithm applied to the graph coloring
  problem}, in: \bibinfo{booktitle}{Proc. 23rd Midwest Artificial Intelligence
  and Cognitive Science Conf}, pp. \bibinfo{pages}{61--66}.
\bibitem[{Jerrum(1995)}]{jerrum1995very}
\bibinfo{author}{Jerrum, M.}, \bibinfo{year}{1995}.
\newblock \bibinfo{title}{A very simple algorithm for estimating the number of
  k-colorings of a low-degree graph}.
\newblock \bibinfo{journal}{Random Structures \& Algorithms}
  \bibinfo{volume}{7}, \bibinfo{pages}{157--165}.
\bibitem[{Kassir(2018)}]{kassir2018absorbing}
\bibinfo{author}{Kassir, A.}, \bibinfo{year}{2018}.
\newblock \bibinfo{title}{Absorbing Markov chains with random transition
  matrices and applications}.
\newblock Ph.D. thesis. University of California, Irvine.
\bibitem[{Kearns et~al.(2006)Kearns, Suri and
  Montfort}]{kearns2006experimental}
\bibinfo{author}{Kearns, M.}, \bibinfo{author}{Suri, S.},
  \bibinfo{author}{Montfort, N.}, \bibinfo{year}{2006}.
\newblock \bibinfo{title}{An experimental study of the coloring problem on
  human subject networks}.
\newblock \bibinfo{journal}{science} \bibinfo{volume}{313},
  \bibinfo{pages}{824--827}.
\bibitem[{Keeney and Kirkwood(1975)}]{keeney1975group}
\bibinfo{author}{Keeney, R.L.}, \bibinfo{author}{Kirkwood, C.W.},
  \bibinfo{year}{1975}.
\newblock \bibinfo{title}{Group decision making using cardinal social welfare
  functions}.
\newblock \bibinfo{journal}{Management Science} \bibinfo{volume}{22},
  \bibinfo{pages}{430--437}.
\bibitem[{Kirkpatrick et~al.(1983)Kirkpatrick, Gelatt~Jr and
  Vecchi}]{kirkpatrick1983optimization}
\bibinfo{author}{Kirkpatrick, S.}, \bibinfo{author}{Gelatt~Jr, C.D.},
  \bibinfo{author}{Vecchi, M.P.}, \bibinfo{year}{1983}.
\newblock \bibinfo{title}{Optimization by simulated annealing}.
\newblock \bibinfo{journal}{science} \bibinfo{volume}{220},
  \bibinfo{pages}{671--680}.
\bibitem[{Kliemann et~al.(2015)Kliemann, Sheykhdarabadi and
  Srivastav}]{kliemann2015price}
\bibinfo{author}{Kliemann, L.}, \bibinfo{author}{Sheykhdarabadi, E.S.},
  \bibinfo{author}{Srivastav, A.}, \bibinfo{year}{2015}.
\newblock \bibinfo{title}{Price of anarchy for graph coloring games with
  concave payoff}.
\newblock \bibinfo{journal}{arXiv preprint arXiv:1507.08249} .
\bibitem[{Lim and Wang(2005)}]{lim2005robust}
\bibinfo{author}{Lim, A.}, \bibinfo{author}{Wang, F.}, \bibinfo{year}{2005}.
\newblock \bibinfo{title}{Robust graph coloring for uncertain supply chain
  management}, in: \bibinfo{booktitle}{Proceedings of the 38th Annual Hawaii
  International Conference on System Sciences}, \bibinfo{organization}{IEEE}.
  pp. \bibinfo{pages}{81b--81b}.
\bibitem[{Luengo and Martino(2013)}]{luengo2013fully}
\bibinfo{author}{Luengo, D.}, \bibinfo{author}{Martino, L.},
  \bibinfo{year}{2013}.
\newblock \bibinfo{title}{Fully adaptive gaussian mixture metropolis-hastings
  algorithm}, in: \bibinfo{booktitle}{2013 IEEE International Conference on
  Acoustics, Speech and Signal Processing}, \bibinfo{organization}{IEEE}. pp.
  \bibinfo{pages}{6148--6152}.
\bibitem[{Marappan and Sethumadhavan(2013)}]{marappan2013new}
\bibinfo{author}{Marappan, R.}, \bibinfo{author}{Sethumadhavan, G.},
  \bibinfo{year}{2013}.
\newblock \bibinfo{title}{A new genetic algorithm for graph coloring}, in:
  \bibinfo{booktitle}{2013 Fifth International Conference on Computational
  Intelligence, Modelling and Simulation}, \bibinfo{organization}{IEEE}. pp.
  \bibinfo{pages}{49--54}.
\bibitem[{Marappan and Sethumadhavan(2021)}]{marappan2021solving}
\bibinfo{author}{Marappan, R.}, \bibinfo{author}{Sethumadhavan, G.},
  \bibinfo{year}{2021}.
\newblock \bibinfo{title}{Solving graph coloring problem using divide and
  conquer-based turbulent particle swarm optimization}.
\newblock \bibinfo{journal}{Arabian Journal for Science and Engineering} ,
  \bibinfo{pages}{1--18}.
\bibitem[{Marden and Shamma(2012)}]{marden2012revisiting}
\bibinfo{author}{Marden, J.R.}, \bibinfo{author}{Shamma, J.S.},
  \bibinfo{year}{2012}.
\newblock \bibinfo{title}{Revisiting log-linear learning: Asynchrony,
  completeness and payoff-based implementation}.
\newblock \bibinfo{journal}{Games and Economic Behavior} \bibinfo{volume}{75},
  \bibinfo{pages}{788--808}.
\bibitem[{Marden and Wierman(2013)}]{marden2013overcoming}
\bibinfo{author}{Marden, J.R.}, \bibinfo{author}{Wierman, A.},
  \bibinfo{year}{2013}.
\newblock \bibinfo{title}{Overcoming the limitations of utility design for
  multiagent systems}.
\newblock \bibinfo{journal}{IEEE Transactions on Automatic Control}
  \bibinfo{volume}{58}, \bibinfo{pages}{1402--1415}.
\bibitem[{Marnissi et~al.(2020)Marnissi, Chouzenoux, Benazza-Benyahia and
  Pesquet}]{marnissi2020majorize}
\bibinfo{author}{Marnissi, Y.}, \bibinfo{author}{Chouzenoux, E.},
  \bibinfo{author}{Benazza-Benyahia, A.}, \bibinfo{author}{Pesquet, J.C.},
  \bibinfo{year}{2020}.
\newblock \bibinfo{title}{Majorize--minimize adapted metropolis--hastings
  algorithm}.
\newblock \bibinfo{journal}{IEEE Transactions on Signal Processing}
  \bibinfo{volume}{68}, \bibinfo{pages}{2356--2369}.
\bibitem[{McQuarrie(2000)}]{mcquarrie2000statistical}
\bibinfo{author}{McQuarrie, D.A.}, \bibinfo{year}{2000}.
\newblock \bibinfo{title}{Statistical mechanics}.
\newblock \bibinfo{publisher}{Sterling Publishing Company}.
\bibitem[{Metropolis et~al.(1953)Metropolis, Rosenbluth, Rosenbluth, Teller and
  Teller}]{metropolis1953equation}
\bibinfo{author}{Metropolis, N.}, \bibinfo{author}{Rosenbluth, A.W.},
  \bibinfo{author}{Rosenbluth, M.N.}, \bibinfo{author}{Teller, A.H.},
  \bibinfo{author}{Teller, E.}, \bibinfo{year}{1953}.
\newblock \bibinfo{title}{Equation of state calculations by fast computing
  machines}.
\newblock \bibinfo{journal}{The journal of chemical physics}
  \bibinfo{volume}{21}, \bibinfo{pages}{1087--1092}.
\bibitem[{Moayedikia et~al.(2020)Moayedikia, Ghaderi and
  Yeoh}]{moayedikia2020optimizing}
\bibinfo{author}{Moayedikia, A.}, \bibinfo{author}{Ghaderi, H.},
  \bibinfo{author}{Yeoh, W.}, \bibinfo{year}{2020}.
\newblock \bibinfo{title}{Optimizing microtask assignment on crowdsourcing
  platforms using markov chain monte carlo}.
\newblock \bibinfo{journal}{Decision Support Systems} \bibinfo{volume}{139},
  \bibinfo{pages}{113404}.
\bibitem[{Monderer and Shapley(1996)}]{monderer1996potential}
\bibinfo{author}{Monderer, D.}, \bibinfo{author}{Shapley, L.S.},
  \bibinfo{year}{1996}.
\newblock \bibinfo{title}{Potential games}.
\newblock \bibinfo{journal}{Games and economic behavior} \bibinfo{volume}{14},
  \bibinfo{pages}{124--143}.
\bibitem[{Panagopoulou and Spirakis(2008)}]{panagopoulou2008game}
\bibinfo{author}{Panagopoulou, P.N.}, \bibinfo{author}{Spirakis, P.G.},
  \bibinfo{year}{2008}.
\newblock \bibinfo{title}{A game theoretic approach for efficient graph
  coloring}, in: \bibinfo{booktitle}{Algorithms and Computation: 19th
  International Symposium, ISAAC 2008, Gold Coast, Australia, December 15-17,
  2008. Proceedings 19}, \bibinfo{organization}{Springer}. pp.
  \bibinfo{pages}{183--195}.
\bibitem[{Pelekis and Schauer(2013)}]{pelekis2013network}
\bibinfo{author}{Pelekis, C.}, \bibinfo{author}{Schauer, M.},
  \bibinfo{year}{2013}.
\newblock \bibinfo{title}{Network coloring and colored coin games}, in:
  \bibinfo{booktitle}{Search Theory: A Game Theoretic Perspective}.
  \bibinfo{publisher}{Springer}, pp. \bibinfo{pages}{59--73}.
\bibitem[{Pickem et~al.(2015)Pickem, Egerstedt and Shamma}]{pickem2015game}
\bibinfo{author}{Pickem, D.}, \bibinfo{author}{Egerstedt, M.},
  \bibinfo{author}{Shamma, J.S.}, \bibinfo{year}{2015}.
\newblock \bibinfo{title}{A game-theoretic formulation of the homogeneous
  self-reconfiguration problem}, in: \bibinfo{booktitle}{2015 54th IEEE
  Conference on Decision and Control (CDC)}, \bibinfo{organization}{IEEE}. pp.
  \bibinfo{pages}{2829--2834}.
\bibitem[{Rosenthal(2006)}]{rosenthal2006first}
\bibinfo{author}{Rosenthal, J.S.}, \bibinfo{year}{2006}.
\newblock \bibinfo{title}{First Look At Rigorous Probability Theory, A}.
\newblock \bibinfo{publisher}{World Scientific Publishing Company}.
\bibitem[{Salari and Eshghi(2005)}]{salari2005aco}
\bibinfo{author}{Salari, E.}, \bibinfo{author}{Eshghi, K.},
  \bibinfo{year}{2005}.
\newblock \bibinfo{title}{An aco algorithm for graph coloring problem}, in:
  \bibinfo{booktitle}{2005 ICSC Congress on computational intelligence methods
  and applications}, \bibinfo{organization}{IEEE}. pp. \bibinfo{pages}{5--pp}.
\bibitem[{Touhiduzzaman et~al.(2018)Touhiduzzaman, Hahn and
  Srivastava}]{touhiduzzaman2018diversity}
\bibinfo{author}{Touhiduzzaman, M.}, \bibinfo{author}{Hahn, A.},
  \bibinfo{author}{Srivastava, A.K.}, \bibinfo{year}{2018}.
\newblock \bibinfo{title}{A diversity-based substation cyber defense strategy
  utilizing coloring games}.
\newblock \bibinfo{journal}{IEEE Transactions on Smart Grid}
  \bibinfo{volume}{10}, \bibinfo{pages}{5405--5415}.
\bibitem[{Vigoda(2000)}]{vigoda2000improved}
\bibinfo{author}{Vigoda, E.}, \bibinfo{year}{2000}.
\newblock \bibinfo{title}{Improved bounds for sampling colorings}.
\newblock \bibinfo{journal}{Journal of Mathematical Physics}
  \bibinfo{volume}{41}, \bibinfo{pages}{1555--1569}.
\bibitem[{Vu et~al.(2014)Vu, Vo and Evans}]{vu2014particle}
\bibinfo{author}{Vu, T.}, \bibinfo{author}{Vo, B.N.}, \bibinfo{author}{Evans,
  R.}, \bibinfo{year}{2014}.
\newblock \bibinfo{title}{A particle marginal metropolis-hastings multi-target
  tracker}.
\newblock \bibinfo{journal}{IEEE transactions on signal processing}
  \bibinfo{volume}{62}, \bibinfo{pages}{3953--3964}.
\bibitem[{Wang and Xu(2013)}]{wang2013metaheuristics}
\bibinfo{author}{Wang, F.}, \bibinfo{author}{Xu, Z.}, \bibinfo{year}{2013}.
\newblock \bibinfo{title}{Metaheuristics for robust graph coloring}.
\newblock \bibinfo{journal}{Journal of Heuristics} \bibinfo{volume}{19},
  \bibinfo{pages}{529--548}.
\bibitem[{Y{\'a}{\~n}ez and Ram{\i}́rez(2003)}]{yanez2003robust}
\bibinfo{author}{Y{\'a}{\~n}ez, J.}, \bibinfo{author}{Ram{\i}́rez, J.},
  \bibinfo{year}{2003}.
\newblock \bibinfo{title}{The robust coloring problem}.
\newblock \bibinfo{journal}{European Journal of Operational Research}
  \bibinfo{volume}{148}, \bibinfo{pages}{546--558}.
\bibitem[{Young(1993)}]{young1993evolution}
\bibinfo{author}{Young, H.P.}, \bibinfo{year}{1993}.
\newblock \bibinfo{title}{The evolution of conventions}.
\newblock \bibinfo{journal}{Econometrica: Journal of the Econometric Society} ,
  \bibinfo{pages}{57--84}.

\end{thebibliography}

\newpage

\appendix

\section{Tabu search algorthm used in Section~\ref{sec: simulation}}
\label{sec: tabu}
\begin{algorithm}[h]
	\KwIn{$G(V, E)$, $M$, $c(0)$, $T$, MaxTenure}
	\KwOut{$c^*_{RWO}$}
	\nl Initialize $c \leftarrow c(0)$, $t \leftarrow 0$, $t' \leftarrow 0$, $tabu_v = [0]^{|V|}$, $tabu_p = [0]^{|V| \times |M|}$.\\
	\nl \While{$t < T$}{
		\nl $t' \leftarrow t' + 1$.\\
		\nl $c' \leftarrow c$.\\
		\nl $\delta \leftarrow 0$.\\
		\nl \For{$V_i \in V$ where $ t' > tabu_v[V_i]$}{
			\nl $c_i^* \leftarrow \argmax_{c_i \in \CC_i = M / c_{\CN(V_i)}} \Phi((c_i, c_{-i}) - \Phi(c) + \expectation(\FL^c) - \expectation(\FL^{(c_i, c_{-i})}))$.\\
			\nl $\delta^* \leftarrow \Phi((c_i^*, c_{-i}) - \Phi(c) + \expectation(\FL^c) - \expectation(\FL^{(c_i^*, c_{-i})}))$.\\
			\nl \If{$\delta^*> \delta$}{
				\nl $V' \leftarrow V_i$; $c^* \leftarrow c_i^*$.\\
				\nl $\delta \leftarrow \delta^*$.\\
			}
		} 
		\nl $c'_{V'} \leftarrow c^*$.\\
		\nl \If{$\delta > 0$}{
			\nl $c \leftarrow c'$.\\
			\nl $t \leftarrow 0$.\\
			\nl $tabu_v[V'] \leftarrow t' + \text{randint}(1, \text{MaxTenure})$.\\
			\nl $tabu_p[V', c^*] \leftarrow t' +~\text{randint}(3, \text{MaxTenure})$.\\
		}\Else{
			\nl $t \leftarrow t + 1$.\\
		}
	}	
	\nl $c^*_{RWO} \leftarrow c$.\\
	\nl \Return $c^*_{RWO}$. \\
	\caption{{\bf Tabu Search Algorithm for \ref{prob: RWO} in asynchronous G-VCGs (Adapted from \cite{lim2005robust})}}
	\label{algo: tabu_search}
\end{algorithm}

\section{Proof of results in Section~\ref{sec: IP-VCG}}
\label{apx:proof_IP-VCG}
\vspace{10pt}

\subsection{Proof of results in Section~\ref{subsec: convergence IP-VCG}}
\label{sapx: convergence IP-VCG}

In this part of the Appendix, we present proofs regarding the lemmas and theorems on the convergence of IP-VCGs, as discussed in Section~\ref{subsec: convergence IP-VCG}.

\begin{proofof}{Lemma~\ref{lemma:AMC}}
Under Assumption~\ref{assp: colors}, the state space $S$ must contain the state $K = (1, 1, \dots, 1)$ corresponding to any proper coloring. One may easily observe that the state $K = (1, 1, \dots, 1)$ is absorbing since none of the agents would accept new colors thus the utility remains. All other states are transient because they are likely to alter in later rounds. It then suffices to prove the accessibility between other states to the state $K$. 

For an active zero-utility agent, say $V_i$, consider the probability $p_i$ of a utility rise from round $t$ to round $t + 1$ . Let $d$ denote the number of $V_i$'s neighbors and $\tilde{d}$ denote the number of colors held by his neighbors in round $t$. We also consider the number of neighbors of $V_i$'s neighbors, denoted by $d_1$, $\cdots$, $d_d$ in respective. Then $V_i$ can only have a utility rise when his new color is unused by his neighbors in round $t$ and does not clash with the new colors of his active neighbors. For a completely synchronous game under Assumption~\ref{assp: colors}, we have
\begin{align*}
	\begin{split}
		p_i & > \frac{m - \tilde{d}}{m} (1 - \frac{1}{m})^{|\{j: V_j \in \CN(V), u_j(t) = 0\}|} \\
		& \geq \frac{m - \delta}{m}(1 - \frac{1}{m})^\delta\\
		& \geq \frac{1}{m} (\frac{1}{2})^{\delta}\\
		& > 0.\\
	\end{split}
\end{align*}
Note that this constant lower bound for $p_i$ is independent of the network structure and neighborhood behaviours. By the definition of the AMC on the utility space, 
\begin{align*}
	P_{kl} = \prod_{i: L^k_i = 0,  L^l_i = 1}p_i \prod_{j: L^l_j= 0}(1 - p_j).
\end{align*}
As such, the probability of transition from a low-welfare state to a high-welfare state is always positive. Then $K = (1, 1, \dots, 1)$ is accessible from any transcient state, which completes the proof. \qed
\end{proofof}

\begin{proofof}{Theorem~\ref{thm: IP-VCG convergence}}
According to the arguments in Section~\ref{sec: IP-VCG}, it suffices to prove that a synchronous IP-VCG with utility functions \ref{eqn: binary utility} converges to some proper coloring. This is equivalent to show that the AMC defined in Theorem~\ref{lemma:AMC} would eventually be absorbed in $K = (1, 1, \dots, 1)$. 

Suppose, from an initial state $L^0$, the chain requires at least $T_i$ rounds to reach an absorbing state and the corresponding probability is $\CP_i$ ($0 < \CP_i < 1$ because the chain is absorbing). Let $T = \max_{i} T_i $ and $\CP = \min_{i} \CP_i$. Then the probability that the chain does not access to the absorbing state after $kT$ rounds is at most $(1 - \CP)^k$, which converges to 0 when $k \to \infty$ (i.e. $\lim_{t \to \infty} Q^t = 0$ ). Therefore, the chain will be absorbed in an absorbing state, which is uniquely $K = (1, 1, \dots, 1)$ in our case. \qed
\end{proofof}

\subsection{Proof of results in Section~\ref{subsec: stochastic boundedness}}
\label{sapx: stochastic boundedness}

In this part of the Appendix, we present proofs regarding the lemmas and theorems on the stochastic boundedness on the time to convergence of IP-VCGs, as discussed in Section~\ref{subsec: stochastic boundedness}.

\begin{proofof}{Theorem~\ref{thm: const_prob_2rounds}}
Denote by $F_i$ the set of $V_i$'s neighbors who have no clash in round $t$, and define $\tilde{F_i} := \cup_{V_u \in F} \{c_u(t)\}$ the set with cardinality denoted by $f_i$. Additionally, define $\tilde{H_i} := \cup_{V_j \in \CN(V_i)}\{c_j(t)\}$. We also define an indicator variable $Y_k$ for each color $k \in M$ as 
\begin{align*}
	Y_k = \begin{cases}
		1 & \textit{if } c_j(t + 1) \neq k, \forall V_j \in \CN(V_i) \\
		0 & \textit{otherwise} .
	\end{cases}
\end{align*}
Define $Y = \sum_{k \in M} Y_k$, which represents the number of colors unused by $V_i$'s neighbors in round $t + 1$. A color $k$ can belong to one of the three cases:
\begin{enumerate}[label=\normalfont{(case \arabic*)},leftmargin=70pt]
	\item $k \in \tilde{F_i}$. Then $\Pr(Y_k = 1) = 0$ because $c_j(t + 1) = c_j(t)$ for $\forall V_j \in F_i$.
	\item $k \in M\backslash \tilde{H_i}$. Then 
	\begin{align*}
		\Pr(Y_k = 1) = (\frac{m - 1}{m})^{|\{V_j \in \CN(V_i): k \notin \tilde{H_j} \cup \{c_j(t)\}\}|};
	\end{align*}
	i.e. the color $k$ can only be available in round $t + 1$ when it is not taken by the active neighbors with zero utility. 
	\item $k \in \tilde{H_i}\backslash \tilde{F_i}$. Then 
	\begin{align*}
		\Pr(Y_k = 1) = (\frac{m - 1}{m})^{|\{V_j \in \CN(V_i): k \notin \tilde{H_j} \cup \{c_j(t)\}\}|}+ \prod_{V_j \in \CN(V_i): c_j(t) = k} \frac{|M\backslash \tilde{ H_j}|}{m}.
	\end{align*}
	The second term results from the fact that an agent would reject any color which is held by at least one of his neighbors and remain the color in the last round.
\end{enumerate}
Therefore, the expectation of $Y$ can be expressed as 
\begin{align}
	\begin{split}
		\expectation(Y) &= \sum_{k \in M} \Pr(Y_k = 1)\\
		&= \sum_{k \in M \backslash \tilde{H_i}} (\frac{m - 1}{m})^{|\{V_j \in \CN(V_i): k \notin \tilde{H_j} \cup \{c_j(t)\}\}|} + \sum_{k \in \tilde{H_i}\backslash \tilde{F_i}} [(\frac{m - 1}{m})^{|\{V_j \in \CN(V_i): k \notin \tilde{H_j} \cup \{c_j(t)\}\}|} + \prod_{V_j \in \CN(V_i): c_j(t) = k} \frac{|M\backslash \tilde{ H_j}|}{m}]\\
		&= \sum_{k \in M \backslash \tilde{F_i}} (\frac{m - 1}{m})^{|\{V_j \in \CN(V_i): k \notin \tilde{H_j} \cup \{c_j(t)\}\}|} + \sum_{k \in \tilde{H_i}\backslash \tilde{F_i}}  \prod_{V_j \in \CN(V_i): c_j(t) = k} \frac{|M\backslash \tilde{ H_j}|}{m} \\
		&\geq \sum_{k \in M \backslash \tilde{F_i}} e^{- \frac{3}{2m}|\{V_j \in \CN(V_i): k \notin \tilde{H_j} \cup \{c_j(t)\}\}|}.
	\end{split}
	\label{eqn: Y-expectation}
\end{align}
Assumption~\ref{assp: colors} implies $m \geq \delta(G) + 1 \geq 2$ because the graph $G$ is connected. The last inequality in \eqref{eqn: Y-expectation} arises from the fact that $1 - x \geq e^{-\frac{2}{3}x}$ for $x \in [0, \frac{1}{2}]$ and the nonnegativity of $\sum_{k \in \tilde{H_i}\backslash \tilde{F_i}}  \prod_{V_j \in \CN(V_i): c_j(t) = k} \frac{|M\backslash \tilde{ H_j}|}{m}$. Defining 
\begin{align*}
	\hat{Y_k} = \begin{cases}
		1 & \textit{if } c_j(t + 2) \neq k, \forall V_j \in \CN(V_i) \\
		0 & \textit{otherwise} 
	\end{cases},
\end{align*}
Lemma~4 and Lemma~5 in \cite{chaudhuri2008network} then apply. Therefore, we can adapt from their Lemma~3 that
\begin{align*}
	\Pr [u_i(t + 2) > 0 | u_i(t) = 0] \geq \BL, \quad \forall t \in \N
\end{align*}
for $\BL = \frac{1}{44100e^{18}}$ by multiplying the original constant by another $\frac{1}{6}$ and $\frac{1}{7e^9}$ due to Assumption~\ref{assp: random1} where $\CC_i = M$. \qed
\end{proofof}

To prepare for the proofs regarding Lemma~\ref{lemma: expectation-variance}, we introduce the canonical form of an AMC.
\begin{definition}[Canonical Form of AMC]
	Suppose an AMC has $r$ absorbing states and $t$ transient states. The transition matrix $P$ of the AMC is in canonical form if it is divided into four sub-matrices listed as follows:
	\begin{align}
		P = 
		\begin{pmatrix}
			Q & R \\
			0 & I \\
		\end{pmatrix}
		\label{def:canonical}
		\tag{CF}
	\end{align}
	where 
	\begin{itemize}
		\item $Q$ is a $t \times t$ block containing the transition probabilities between transient states.
		\item $R$ is a $t \times r$ block containing the transition probabilities from transient to absorbing states.
		\item $I$ is the identity matrix of order $r$ and $0$ is the block with null elements
	\end{itemize}
\end{definition}
Since an AMC will be eventually absorbed in an absorbing state, as we prove in Theorem~\ref{thm: IP-VCG convergence}, we have
\begin{align*}
	\lim_{t \rightarrow \infty} Q^t = 0.
\end{align*}
The following lemma gives several properties of an AMC which would be useful in our later arguments.
\begin{lemma}[Properties of AMC (See, e.g. \cite{kassir2018absorbing} Chapter 2)]
	For an AMC with transition matrix $P$ in canonical form \eqref{def:canonical}, then
	\begin{enumerate}[label=\normalfont{(AMC-\arabic*)},leftmargin=50pt]
		\item \label{amc1}$(I - Q)^{-1}$ exists.
		\item \label{amc2}$\lim_{t \to \infty}P^t = $
		$
		\begin{pmatrix}
			0 & (I - Q)^{-1}R\\
			0 & I\\
		\end{pmatrix}
		$
		\item \label{amc3}Suppose the number of states is $k$. Define a $k \times 1$ column vector $n$ whose i-th element $\Bn_i$ denotes the expected number of steps before absorption, given the initial state $i$. Then $\Bn = (I - Q)^{-1} \mathds{1}$ where $\mathds{1} = (1, 1, \cdots, 1)_k^ \intercal$.
	\end{enumerate}
	\label{lemma: AMC properties}
\end{lemma}

\begin{proofof}{Lemma~\ref{lemma: AMC properties}}
The first statement is equivalent to``if $(I - Q)x = 0$, then $x = 0$" which is easy to prove as
\begin{align*}
	(I - Q)x = 0 \iff x = Qx 
	\iff x = \lim_{t \to \infty}Q^t x
	\iff x = 0.
\end{align*}

Let $N = (I - Q)^{-1}$, then $N = \sum_{t = 0}^{\infty}Q^t$ by Taylor's formula. Notice that the n-th power of the transition matrix is
\begin{align*}
	P^t =
	\begin{pmatrix}
		Q^t & (I + Q + \cdots + Q^{t - 1})R \\
		0 & I\\
	\end{pmatrix}
\end{align*}
then
\begin{align*}
	\lim_{t \to \infty}P^t = 
	\begin{pmatrix}
		0 & (\sum_{t = 0}^{\infty}Q^t)R)\\
		0 & I\\
	\end{pmatrix}
	= \begin{pmatrix}
		0 & NR\\
		0 & I\\
	\end{pmatrix}.
\end{align*}

Denote by $X_{ij}$ the times the chain hits state $j$ starting from $i$ and $X_{ij}^{(l)}$ the indicator of arriving at state $j$ in the $l$th step, where $i$ and $j$ are both transient states. Then
\begin{align*}
	\begin{split}
		\mathds{E}(X_{ij}) & = \sum_{l = 0}^{\infty} \Pr{(X_{ij}^{(l)} = 1)} \\
		& = I_{ij} + Q_{ij} + Q_{ij}^2 + Q_{ij}^3 + \cdots\\
		& = N_{ij}.
	\end{split}
\end{align*}
Since $\Bn_i := \mathds{E}(X_i)$ where $X_i$ denotes the number of steps before absorption given the initial state $i$, we have
\begin{align*}
	\begin{split}
		\Bn_i & = \mathds{E}(\sum_{j = 1}^{k} X_{ij}) = \sum_{j = 1}^{k} N_{ij} = [N \mathds{1}]_i
	\end{split}
\end{align*}
which completes the proof. \qed
\end{proofof}

Next, we prove Lemma~\ref{lemma: expectation-variance}.
\begin{proofof}{Lemma~\ref{lemma: expectation-variance}}
It suffices to prove the lemma on the representative candidate of IP-VCGs with binary utility functions \eqref{eqn: binary utility}, as explained in Section~\ref{sec: IP-VCG}. Given an arbitrary number $\epsilon \in (0, 1)$ and define $\tau =  \frac{1}{\BL} \ln (\frac{n}{\epsilon})$. We also define the random variable $T$ to denote the time to convergence of the game, and, specifically, let $T_i$ denote the number of steps to convergence of $(X_t)_{t \in \N}$ initiating from the state $i$. Invoking Theorem~\ref{thm: const_prob_2rounds} we have
\begin{align}
	\begin{split}
		\Pr[u_i(c(2\tau)) = 0 | u_i(c(0)) = 0] & = \prod_{t = 1}^{\tau} \Pr[u_i(c(2t)) = 0 | \bigcap_{s = 1}^{t - 1}  \Pr[u_i(c(2s)) = 0]
		\\ &\leq (1 - \BL)^\tau 
		\\ &\leq e^{-\BL \tau}.
	\end{split}
	\label{eqn: 111}
\end{align}
The first equality in \eqref{eqn: 111} is due to the fact that agents with utility one will remain his color forever. Thus,
\begin{align*}
	\Pr[u_i(c(2\tau)) = 1 | u_i(c(0)) = 0] \geq 1 - \frac{\epsilon}{n}.
\end{align*}
Taking an union bound over all vertices, we obtain 
\begin{align*}
	\Pr[T \leq 2\tau] & \geq (1 - \frac{\epsilon}{n})^n \geq 1 - \epsilon
\end{align*}
thus 
\begin{align}
	\Pr[T > 2\tau] \leq \epsilon.
	\label{eqn: T-bound}
\end{align}

We generate the Markov Chain defined in Lemma~\ref{lemma:AMC}, $(X_t)_{t \in \N}$, to jump two steps at a time and consider its transition matrix $P$ in the form \eqref{def:canonical}. Define $\tilde{N} := (I - Q^2)^{-1}$. We denote $\Bn = (\Bn_1, \Bn_2, \dots, \Bn_{2^n - n - 1})^\TRANSP$ the column vector whose i-th element $\Bn_i$ denotes the expected number of steps before absorption starting from $i \neq K = (1, 1, \dots, 1)$, i.e. $\expectation [T_i]$, where we correspond $K$ to the $2^n- n$ row in $P$. Obviously $\Bn_{2^n - n} = 0$ corresponding to the absorbing state of any proper coloring. By  \ref{amc3} of Lemma~\ref{lemma: AMC properties},
\begin{align*}
	2 \tilde{N} \indicator \geq  \Bn
\end{align*}
elementwise. For each element $n_i$, equivalently,
\begin{align*}
	\begin{split}
		\Bn_i & \leq 2 \sum_{\tau = 0}^{\infty} \sum_{j = 1}^{2^n - n - 1} {Q^2}_{ij}^{(\tau)} = 2 \sum_{\tau = 0}^{\infty} \Pr[T > 2\tau].
	\end{split}
\end{align*}

Along with \eqref{eqn: T-bound} for $\epsilon = ne^{-\BL \tau}$, we have
\begin{align}
	\begin{split}
		\sum_{\tau = 0}^{\infty} \Pr[T > 2\tau] &\leq 1 \times \lceil \frac{\ln{n}}{\BL} \rceil + \sum_{\tau = \lceil \frac{\ln{n}}{\BL} \rceil}^{\infty} ne^{-\BL \tau}
		\\ & = \lceil \frac{\ln{n}}{\BL} \rceil + n\sum_{\tau = \lceil \frac{\ln{n}}{\BL} \rceil}^{\infty} e^{-\BL \tau}
		\\ & \leq \lceil \frac{\ln{n}}{\BL} \rceil + n(\frac{1}{n} \frac{1}{1 - e^{-\BL}})
		\\ & = \CO \log n.
	\end{split}
\end{align}
Therefore, $\expectation[T] = \CO \log n$.

Denote $\Br = (\Br_1, \Br_2, \dots, \Br_{2^n - n - 1})^\TRANSP$ be the column vector whose i-th element $\Br_i$ denotes $\variance[T_i]$, the variance of steps before absorption starting from $i \neq K = (1, 1, \dots, 1)$. Then
\begin{align*}
	\Br_i = \expectation [T_i^2] - \Bn_i^2.
\end{align*}
We consider all possible states led by the first step of $X_{ij}^{(l)}$ and derive an expression for $\expectation[T_i]$ as a probabilistic combination of the expected number of steps given other initiate states. 
\begin{align}
	\begin{split}
		\mathds{E}[T_i^2] &= \sum_{k = 1}^{2^n - n} P_{ik} \expectation[(1+T_k)^2]
		\\ & = \sum_{k = 1}^{2^n - n} P_{ik} \expectation[T_k^2 + 2T_k + 1]
		\\ & = 1 + 2\sum_{k = 1}^{2^n - n} P_{ik} \expectation[T_k] + \sum_{k = 1}^{2^n - n} P_{ik} \mathds{E}[T_k^2]
	\end{split}
	\label{eqn: first step analysis}.
\end{align}
This operation is often named ``first-step analysis'' in stochastic processes. 

Let $\expectation[T^2] = (\expectation[T_1^2], \expectation[T_2^2], \dots, \expectation[T_{2^n - n - 1}^2])^\TRANSP$. We rewrite \eqref{eqn: first step analysis} in vector form as 
\begin{align*}
	\mathds{E} [T^2] = \indicator + 2Q\Bn + Q \expectation [T^2]
\end{align*}
and simple arrangements give
\begin{align*}
	\begin{split}
		\expectation [T^2]  & = (I - Q)^{-1}(\indicator + 2Q \Bn)
		\\ & = \Bn+ 2(I - Q)^{-1}Q \Bn
		\\ & = \Bn + 2(Q + Q^2 + \dots) \Bn
		\\ & = \Bn + 2[(I - Q)^{-1}- I] \Bn
		\\ & = [2\Bn- I] \Bn
		\\ & = \CO (\log n)^2.
	\end{split}
\end{align*}
Since $(\expectation[T])^2$ is also $\CO (\log n)^2$, we know that $\variance[T_i] = \CO (\log n)^2$. 

The proof is thus complete. \qed
\end{proofof}

We now forward to prove Theorem~\ref{thm:stochastic}.
\begin{proofof}{Theorem~\ref{thm:stochastic}}
Let $\mu^{(n)} = \expectation[T^{(n)}]$ and $\sigma^{(n)} = \sqrt{\variance [T^{(n)}]}$. By Lemma~\ref{lemma: expectation-variance}, $\mu^{(n)} \leq D_1 \log n$ for $n \geq N_1$ and $\sigma^{(n)} \leq D_2 \log n$ for $n \geq N_2$, where $D_1$ and $D_2$ are two positive constants. By Chebyshev's inequality, $\forall k \geq 1$,
\begin{align*}
	\Pr [T^{(n)} - \mu^{(n)} \geq k \sigma^{(n)}] \leq \Pr [|T^{(n)} - \mu^{(n)}| \geq k \sigma^{(n)}] \leq \frac{1}{k^2}.
\end{align*}
For $\forall \epsilon > 0$, let $M_\epsilon = \frac{D_2}{\sqrt{\epsilon}} + D_1 > D_1$. Plugging $k = \frac{M_\epsilon \log n - \mu^{(n)}}{\sigma^{(n)}}$ into the inequality gives
\begin{align*}
	\Pr [T^{(n)} - \mu^{(n)} > M_\epsilon \log n - \mu^{(n)}] < \frac{D_2^2(\log n)^2}{(M_\epsilon \log n - \mu^{(n)})^2}
\end{align*}
thus 
\begin{align*}
	\Pr[\frac{T^{(n)}}{\log n} > M_\epsilon] < \epsilon, \quad \forall n > N_\epsilon = \max\{N_1, N_2\}
\end{align*}
which completes the proof. \qed
\end{proofof}

\section{Proof of results in Section~\ref{sec: MH-policy}}
\label{apx: MH-policy}

\subsection{Proof of results in Section~\ref{subsec: MH-policy}}
\label{sapx: MH-policy}
In this part of the Appendix, we present proofs regarding the lemmas and theorems on the the optimality of the proposed MH-policy, as discussed in Section~\ref{subsec: MH-policy}.

\begin{proofof} {Lemma~\ref{lemma: MH-induced RPMP}}
We prove by definition. Irreducibility follows from the proposal uniform distribution due to  ramdom sampling from $\CC_i = M$ for $\forall V_i \in V$ being active. The transition probability from state (coloring) $c$ to state (coloring) $c'$ by the MH-policy (Algorithm~\ref{algo: MH-policy}) is 
\begin{align*}
	P^\epsilon_{c \rightarrow c'}  = \frac{1}{n} \frac{1}{m} \min \{1, \epsilon^{-\Delta U_{\CF_i \cup \{V_i\}} (c \rightarrow c')}\}
\end{align*}
where $\epsilon = e^{-\frac{1}{\tau}}$. Note that $P^\epsilon_{c \rightarrow c'} > 0$ whatever $c$ and $c'$ are when $\epsilon < \infty$. Then we have
\begin{align*}
	\lim_{\epsilon \rightarrow 0^+} P^\epsilon_{c \rightarrow c'} = \frac{1}{n} \frac{1}{m} \indicator_{U_{\CF_i \cup \{V_i\}}(c) < U_{\CF_i \cup \{V_i\}}(c')}
\end{align*}
which is the transtion probability of the unperturbed distribution $P^0_{c \rightarrow c'}$. Then we have
\begin{align*}
	\frac{P^\epsilon_{c \rightarrow c'}}{\epsilon^{\max \{0, - \Delta U_{\CF_i\cup \{V_i\}} (c \rightarrow c‘)\}}}  = \frac{1}{nm}> 0
\end{align*}
thus the MH-policy induced Markov Chain is RPMP with resistances
\begin{align}
	R(c \rightarrow c') = \max \{0, - \Delta U_{\CF_i\cup \{V_i\}} (c \rightarrow c')\}.
	\label{eqn: resistance of MH}
\end{align}
\qed
\end{proofof}

The proof of Theorem~\ref{thm: MH-policy optimal} is based on Lemma~\ref{lemma: MH-induced RPMP}.

\begin{proofof} {Theorem~\ref{thm: MH-policy optimal}}
We prove by contradiction. Let $c$ be a stochastically stable state and $T^*_c$ is the corresponding minimum resistance tree rooted at $c$; i.e. $R(T^*_c) = \gamma^* := \min_{j \in \{1, 2, \dots, |C|\}} \gamma_j$ by Theorem~\ref{thm: characterization of stochastic-stable}. Suppose $c \neq \argmax_{\tilde{c} \in C} \Phi(\tilde{c})$. Consider an optimal state $c'$ such that $c' = \arg \max_{\tilde{c} \in C} \Phi(\tilde{c})$. We try to establish a new tree $T_{c'}$ rooted at $c'$ such that $R(T_{c'}) < R(T^*_c)$.

By Definition~\ref{def: resistance tree}, there exists a unique path
\begin{align*}
	\CP(c' \rightarrow c) = \{c' \rightarrow c^1 \rightarrow \dots \rightarrow c^k \rightarrow c \}
\end{align*}
from $c'$ to $c$. Consider the reverse of this path
\begin{align*}
	\CP(c \rightarrow c') = \{c \rightarrow c^k \rightarrow \dots \rightarrow c^1 \rightarrow c' \}
\end{align*}
and one can observe that a new tree, namely $T_{c'}$ can be established by replacing $\CP(c' \rightarrow c)$ with $\CP(c \rightarrow c')$. The vertices which used to access $c$ without passing $c'$ would now access $c'$ uniquely through $\CP(c \rightarrow c')$. Figure~\ref{fig: replace_path} illustrates an example when $|C| = 8$ and $k = 2$ with $T^*_c$ and $T_{c'}$ represented in red and green respectively. The resistence of $T_{c'}$ is then
\begin{align}
	R(T_{c'}) = R(T^*_c) + R(\CP(c \rightarrow c')) - R(\CP(c' \rightarrow c)).
	\label{eqn: two resistance}
\end{align}

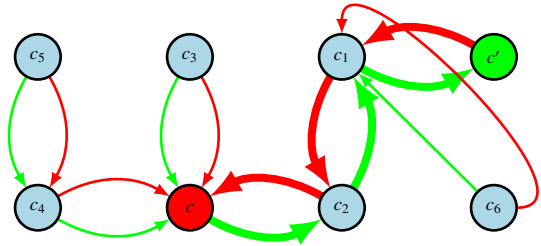
\begin{figure}[h]
	\centering
	\begin{tikzpicture}
		\Vertex [label = $c$, color=red] {s}  \Vertex [x = 2, label = $c_2$]{s2}
		\Vertex [x = 2, y = 2, label = $c_1$]{s1}
		\Vertex [x = 4, y = 2 , label = $c'$, color = green]{s'} \Vertex [x = 4, y = 0, label = $c_6$]{s6}
		\Vertex [y = 2, label = $c_3$]{s3} \Vertex [x = -2, label = $c_4$]{s4}
		\Vertex [x = -2, y = 2, label = $c_5$]{s5}
		\Edge [Direct, lw = 1, bend = 30, color = red](s5)(s4)
		\Edge [Direct, lw = 1, bend = -30, color = green](s5)(s4)
		\Edge [Direct, lw = 1, bend = 30, color = red](s4)(s)
		\Edge [Direct, lw = 1, bend = -30, color = green](s4)(s)
		\Edge [Direct, lw = 1, bend = 30, color = red](s3)(s)
		\Edge [Direct, lw = 1, bend = -30, color = green](s3)(s)
		\Edge [Direct, lw = 3, bend = -30, color = red](s2)(s)
		\Edge [Direct, lw = 3, bend = -30, color = green](s)(s2)
		\Edge [Direct, lw = 3, bend = -30, color = red](s1)(s2)
		\Edge [Direct, lw = 3, bend = -30, color = green](s2)(s1)
		\Edge [Direct, lw = 3, bend = -30, color = red](s')(s1)
		\Edge [Direct, lw = 3, bend = -30, color = green](s1)(s')
		\Edge [Direct, lw = 1, bend = -135, color = red](s6)(s1)
		\Edge [Direct, lw = 1, color = green](s6)(s1)
	\end{tikzpicture}
	\caption{Example: establish $T_{c'}$ from $T^*_c$}
	\label{fig: replace_path}
\end{figure}

By \eqref{eqn: resistance of MH} and \eqref{eqn: welfare-family}, for adjacent states $c^l$ and $c^{l + 1}$ (i.e. there is a an edge between $c^l$ and $c^{l + 1}$),
\begin{align*}
	R(c^l \rightarrow c^{l + 1}) = \max \{0, \Phi(c^l) - \Phi(c^{l + 1})\} \\
	R(c^{l + 1} \rightarrow c^l) = \max \{0, \Phi(c^{l + 1}) - \Phi(c^l)\} \\
\end{align*}
thus 
\begin{align*}
	R(c^{l + 1} \rightarrow c^l) - R(c^l \rightarrow c^{l + 1}) = \Phi(c^{l + 1}) - \Phi(c^l).
\end{align*}
Therefore, with $c^0 = c'$ and $c^{k + 1} = c$,
\begin{align*}
	R(\CP(c \rightarrow c')) - R(\CP(c' \rightarrow c)) &= \sum_{l = 0}^{k} [R(c^l \rightarrow c^{l + 1}) - R(c^{l + 1} \rightarrow c^l)] \\
	& = \Phi(c) - \Phi(c') \\
	& < 0
\end{align*}
because $c'$ achieves best welfare by assumption. Plugging back into \eqref{eqn: two resistance}, we have 
\begin{align*}
	R(T_{c'}) < R(T^*_c)
\end{align*}
which contradicts to our assumption that $c$ is stochastically stable. We now conclude that any stochastically stable state(coloring) must bring the largest welfare; i.e. a G-VCG converges to $c^*_{WO2}$ under the MH-policy. \qed
\end{proofof}

\subsection{Proof of results in Section~\ref{subsec: MH indep_synchronous}}
\label{sapx: MH indep_synchronous}
In this part of the Appendix, we present proofs regarding the lemmas and theorems on the the optimality of the proposed MH-policy in independent synchronous settings, as discussed in Section~\ref{subsec: MH indep_synchronous}.

\begin{proofof} {Lemma~\ref{lemma: MH-induced RPMP_indep-synchronous}}
\ref{def: RPMC1} and \ref{def: RPMC2} in Definition~\ref{def: RPMC} are obviously satisfied with same arguments as in our proof of Lemma~\ref{lemma: MH-induced RPMP}. We now examine whether \ref{def: RPMC3} holds.
\begin{align}
	\begin{split}
		P_t^{\epsilon_t} (c \rightarrow c') &= \sum_{S \subseteq V: G \subseteq S} \omega_t^{|S|} (1 - \omega_t)^{|V \backslash S|} \times \prod_{V_i \in G} \frac{1}{m} \min \{1, \epsilon_t^{-\Delta U_{\CF_i\cup \{V_i\}} ((c_i, c_{-i}) \rightarrow (c_i', c_{-i}))}\} \\
		& \quad \quad \times \prod_{V_j \in S \backslash G} [1 - \sum_{\tilde{c} \in \CC_j \backslash \{c_j\}} \frac{1}{m} \min \{1, \epsilon_t^{-\Delta U_{\CF_j\cup \{V_j\}} ((c_j, c_{-j}) \rightarrow (\tilde{c}_j, c_{-j}))}\}] \\
		& = \sum_{S \subseteq V: G \subseteq S} \omega_t^{|S|} (1 - \omega_t)^{|V \backslash S|}  \frac{1}{m^{|G|}} \epsilon_t ^ {\sum_{V_i \in G} \max \{0, -\Delta U_{\CF_i\cup \{V_i\}} ((c_i, c_{-i}) \rightarrow (c_i', c_{-i}))\}}  \\
		& \quad \quad \times \underbrace{\prod_{V_j \in S \backslash G} [1 - \sum_{\tilde{c} \in M \backslash \{c_j\}} \frac{1}{m} \underbrace{\epsilon_t ^ {\sum_{V_i \in G} \max \{0, -\Delta U_{\CF_i\cup \{V_i\}} ((c_i, c_{-i}) \rightarrow (\tilde{c}_i, c_{-i}))\}}}_{\in (0, 1)}]}_{\in (\frac{1}{m^{|S \backslash G|}}, 1)}\\
	\end{split}
	\label{eqn: transition prob_0}
\end{align}
thus 
\begin{align*}
	\begin{split}
		\lim_{t \rightarrow \infty} \frac{P_t^{\epsilon_t} (c \rightarrow c')}{\epsilon_t ^ {R^{\epsilon_t}(c \rightarrow c')}} &\in (\sum_{S \subseteq V: G \subseteq S} \omega_t^{|S|} (1 - \omega_t)^{|V \backslash S|} \frac{1}{m^{|S|}}, \sum_{S \subseteq V: G \subseteq S} \omega_t^{|S|} (1 - \omega_t)^{|V \backslash S|} \frac{1}{m^{|G|}}) \\
		& \in (0, +\infty)		
	\end{split}
\end{align*}
where
\begin{align*}
	R^{\epsilon_t}(c \rightarrow c') = \sum_{V_i \in G} \max \{0, -\Delta U_{\CF_i\cup \{V_i\}} ((c_i, c_{-i}) \rightarrow (\tilde{c}_i, c_{-i}))\}.
\end{align*}
\qed
\end{proofof}

The proof of Theorem~\ref{thm: optimality MH synchronous} is based on Lemma~\ref{lemma: MH-induced RPMP_indep-synchronous}.

\begin{proofof} {Theorem~\ref{thm: optimality MH synchronous}}
The proof is similar to the one in \cite{marden2012revisiting} Theorem 4.2. Arbitrarily consider a transition $c \rightarrow c'$ where $|G| \equiv \{V_j: c_j \neq c_j'\} \geq 2$ in round $t$. Given the independent activation probability $\omega_t = \epsilon_t^\CK$ where $\epsilon_t = e^{- \frac{1}{\tau(t)}}$, the probability of this transition is 
\begin{align}
	\begin{split}
		P_t^{\epsilon_t} (c \rightarrow c') &=  \epsilon_t^{\CK |G|} \sum_{S \subseteq V: G \subseteq S} \biggl \{\epsilon_t^{\CK(|S| - |G|)} (1 - \epsilon_t^\CK)^{|V \backslash S|} \times \prod_{V_i \in G} \frac{1}{m} \min \{1, \epsilon_t^{-\Delta U_{\CF_i\cup \{V_i\}} ((c_i, c_{-i}) \rightarrow (c_i', c_{-i}))}\} \\
		& \quad \quad \times \prod_{V_j \in S \backslash G} [1 - \sum_{\tilde{c} \in M \backslash \{c_j\}} \frac{1}{m} \min \{1, \epsilon_t^{-\Delta U_{\CF_j\cup \{V_j\}} ((c_j, c_{-j}) \rightarrow (\tilde{c}_j, c_{-j}))}\}] \biggl \}.\\
	\end{split}
	\label{eqn: transition prob}
\end{align}
Note that the second equality of \eqref{eqn: transition prob} results from \ref{assp: random1} in Assumption~\ref{assp: random}. Dividing $P_t^{\epsilon_t} (c \rightarrow c')$ by $\epsilon_t ^ {\CK |G| + \sum_{i \in G} \max \{0, -\Delta U_{\CF_i\cup \{V_i\}} ((c_i, c_{-i}) \rightarrow (c_i', c_{-i}))\}}$ and taking limits $t \rightarrow \infty$ on both sides of \eqref{eqn: transition prob} (equivalently $\tau(t) \rightarrow 0$ and $\epsilon_t \rightarrow 0$) eliminates the terms when $|S| > |G|$ and gives
\begin{align*}
	\begin{split}
		&\lim_{t \rightarrow \infty} \frac{P_t^{\epsilon_t} (c \rightarrow c')}{\epsilon_t ^ {\CK |G| + \sum_{V_i \in G} \max \{0, -\Delta U_{\CF_i\cup \{V_i\}} ((c_i, c_{-i}) \rightarrow (c_i', c_{-i}))\}}} \\
		=&\lim_{t \rightarrow \infty} \frac{\prod_{V_i \in G} \frac{1}{m} \min \{1, \epsilon_t^{-\Delta U_{\CF_i\cup \{V_i\}} ((c_i, c_{-i}) \rightarrow (c_i', c_{-i}))}\}}{\epsilon_t ^ {\sum_{V_i \in G} \max \{0, -\Delta U_{\CF_i\cup \{V_i\}} ((c_i, c_{-i}) \rightarrow (c_i', c_{-i}))\}}}\\
		=&\lim_{t \rightarrow \infty} \frac{1}{m^{|G|}}\frac{\prod_{V_i \in G} \epsilon_t^{\max\{0, -\Delta U_{\CF_i\cup \{V_i\}} ((c_i, c_{-i}) \rightarrow (c_i', c_{-i}))\}}}{\epsilon_t ^ {\sum_{V_i \in G} \max \{0, -\Delta U_{\CF_i\cup \{V_i\}} ((c_i, c_{-i}) \rightarrow (c_i', c_{-i}))\}}} \\
		=& \frac{1}{m^{|G|}} \in (0, +\infty).
	\end{split}	
\end{align*}
Therefore, the process induced by the MH-policy is exactly an RPMP with transition resistance 
\begin{align*}
	R(c \rightarrow c') = \CK |G| + \sum_{V_i \in G} \max \{0, -\Delta U_{\CF_i\cup \{V_i\}} ((c_i, c_{-i}) \rightarrow (c_i', c_{-i}))\}.
\end{align*}

Denote $\phi^*$ to be the maximum preference among all agents and colors. Then for any transition $c \rightarrow c'$ with $G := \{V_j: c_j \neq c_j'\}$, 
\begin{align}
	\CK|G| \leq R(c \rightarrow c') \leq \CK|G| + \phi^* |G|.
	\label{eqn: bound for resistance}
\end{align}
We would like to show that any edge $\{c^k \rightarrow c^l\}$ in a minimum resistance tree $\CT^*_{\tilde{c}}$, whatever the root $\tilde{c}$ is, must have its endpoints deviating only in single element; i.e. $|\{V_j: c_j^k \neq c_j^l\}| = 1$. Then the optimality of a stochastically stable state is covered by our proof in Theorem~\ref{thm: MH-policy optimal}. 

Again we prove by contradiction. Suppose there exists an edge $\{c \rightarrow c'\}$ in $\CT^*_{\tilde{c}}$ such that $c$ and $c'$ have more than one deviators; i.e. $G:= \{V_j: c_j \neq c_j'\}$ has cardinality $|G| \geq 2$. We now look for another path $\CP_{c \rightarrow c'} = \{c \rightarrow c^1 \rightarrow c^2 \rightarrow \dots \rightarrow c^{k - 1} \rightarrow c'\}$ where $k = |G|$ such that the adjacent coloring only differ in one element. By \eqref{eqn: bound for resistance}, 
\begin{align*}
	\begin{split}
		R(c \rightarrow c') &\geq \CK|G| \\
		R(\CP_{c \rightarrow c'}) &= \sum_{i = 0}^{k - 1} R(c^i \rightarrow c^{i + 1}) \leq |G|(\CK + \phi^*)
	\end{split}
\end{align*}
where $c^0 := c$ and $c^k := c'$. Ignoring the original path $\CP_{c^i \rightarrow \tilde{c}}$ for $i = 0, 1, 2, \dots, k - 1$ and implementing $\CP_{c \rightarrow c'}$ into $\CT^*_{\tilde{c}}$ gives another tree $\CT_{\tilde{c}}$. Then 
\begin{align*}
	\begin{split}
		R(\CT_{\tilde{c}}) &= R(\CT^*_{\tilde{c}}) + R(\CP_{c \rightarrow c'}) - R(c \rightarrow c') - \sum_{i = 1}^{k - 1}R(\CP_{c^i \rightarrow \tilde{c}})\\
		&\leq R(\CT^*_{\tilde{c}}) + \CK |G| + \phi^*|G| - \CK |G| - \CK (|G| - 1) \\
		&= R(\CT^*_{\tilde{c}}) + \phi^*|G| - \CK(|G| - 1).
	\end{split}
\end{align*}
As long as $K > \frac{|G| \phi^*}{|G| - 1}$, one has $R(\CT_{\tilde{c}}) < R(\CT_{\tilde{c}^*})$ which contradicts out assumption that $\CT^*_{\tilde{c}}$ has the least resistance. Therefore, a minimum resistance tree must have its endpoints deviating only in single element and the rest follows from Theorem~\ref{thm: MH-policy optimal}.\qed
\end{proofof}

\subsection{Group B: WO-Async 2}
\label{ssec: WO-Async 2}
\begin{figure}[h]
	\centering
	\begin{subfigure}{0.49\textwidth}
		\centering
		\includegraphics[width = \linewidth]{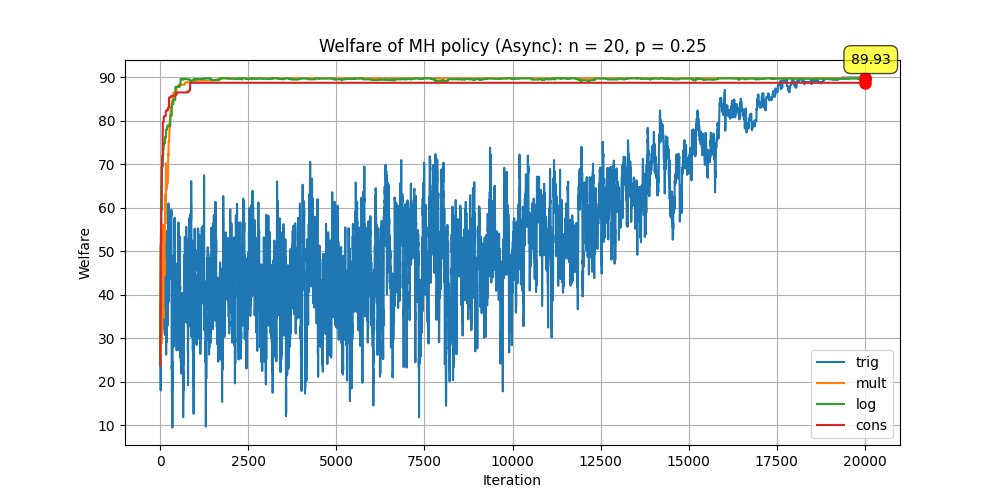}
		\caption{MH-policy: $n = 20$, $p = 0.25$}
		\label{fig: WO_async_0.25}
	\end{subfigure}
	\hfill
	\begin{subfigure}{0.49\textwidth}
		\centering
		\includegraphics[width = \linewidth]{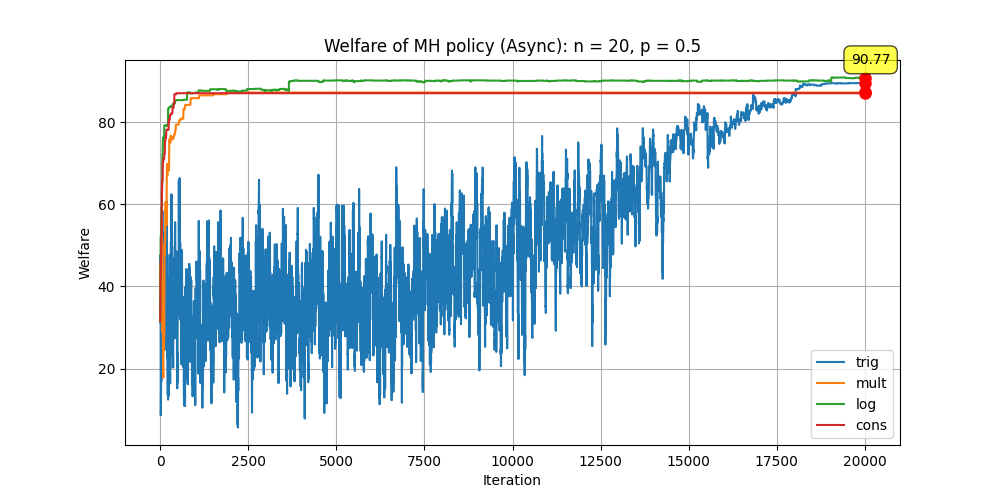}
		\caption{MH-policy: $n = 20$, $p = 0.5$}
		\label{fig: WO_async_0.5}
	\end{subfigure}
	
	\begin{subfigure}{0.49\textwidth}
		\centering
		\includegraphics[width = \linewidth]{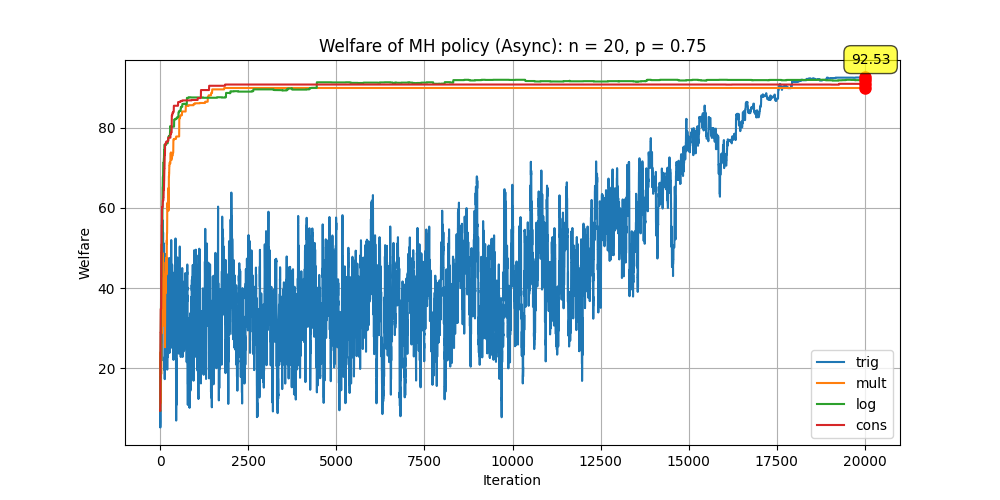}
		\caption{MH-policy: $n = 20$, $p = 0.75$}
		\label{fig: WO_async_0.75}
	\end{subfigure}
	\hfill
	\begin{subfigure}{0.49\textwidth}
		\centering
		\includegraphics[width = \linewidth]{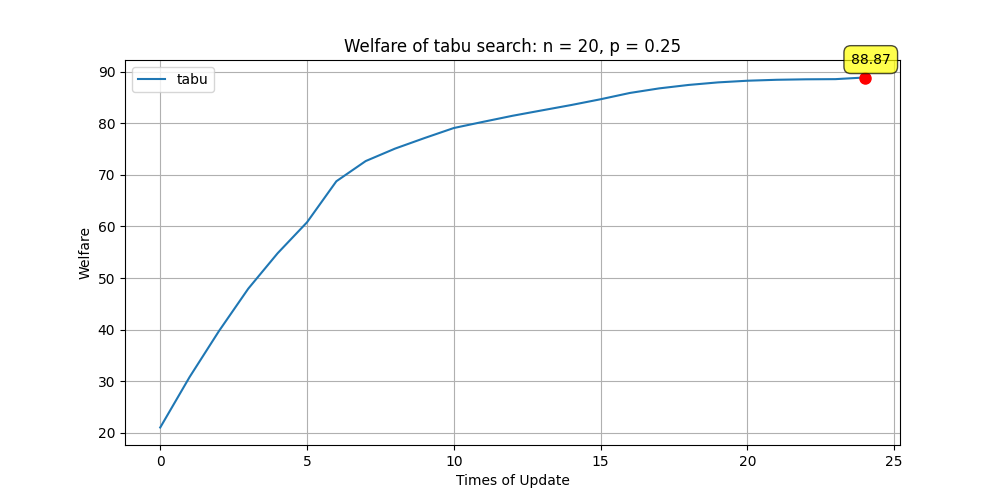}
		\caption{TS: $n = 20$, $p = 0.25$}
		\label{fig: WO_async_TS_0.25}
	\end{subfigure}
	
	\begin{subfigure}{0.49\textwidth}
		\centering
		\includegraphics[width = \linewidth]{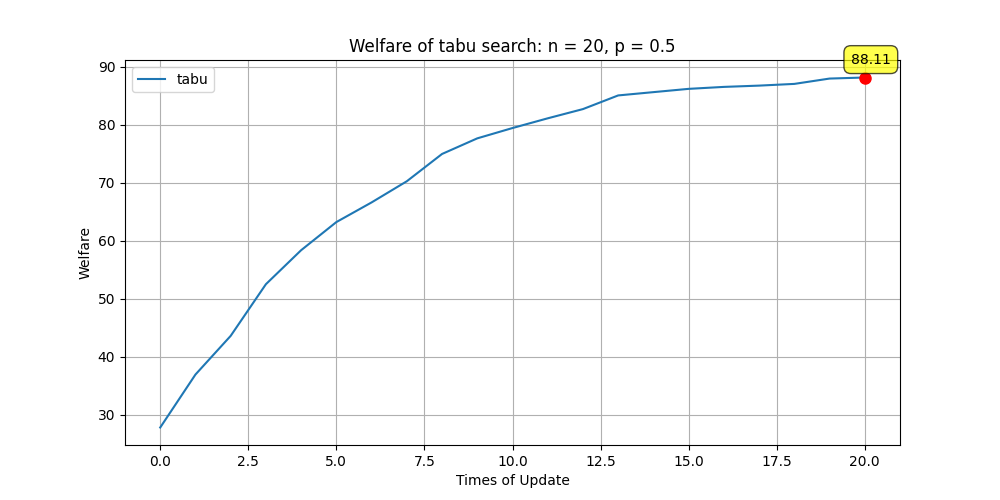}
		\caption{TS: $n = 20$, $p = 0.5$}
		\label{fig: WO_async_TS_0.5}
	\end{subfigure}
	\hfill
	\begin{subfigure}{0.49\textwidth}
		\centering
		\includegraphics[width = \linewidth]{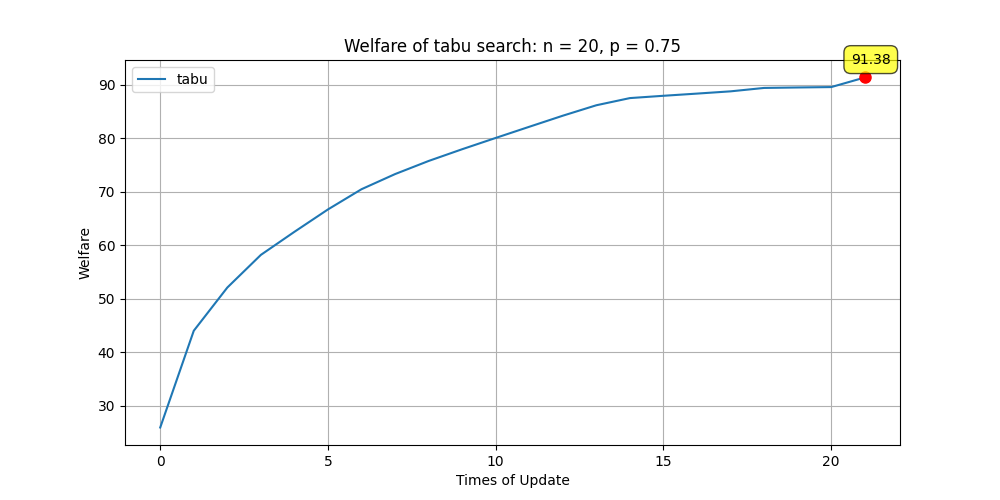}
		\caption{TS: $n = 20$, $p = 0.75$}
		\label{fig: WO_async_TS_0.75}
	\end{subfigure}
	
	\caption{Dynamics of solving \ref{prob: optimalcoloring2} for different network degrees.}
	\label{fig: WO_async_degree}
\end{figure}

In this group, we investigate on the performance of Algorithm~\ref{algo: MH-policy} when given networks imposed with different connection probabilities (thus different degrees) while control the sizes identical ($n = 20$), and compare the results with the TS algorithm~\ref{algo: tabu_search}. See Figure~\ref{fig: WO_async_degree} for the results.

\subsection{Group C: WO-Sync}
\label{ssec: WO-Sync}
\begin{figure}[h]
	\centering
	\begin{subfigure}{0.49\textwidth}
		\centering
		\includegraphics[width = \linewidth]{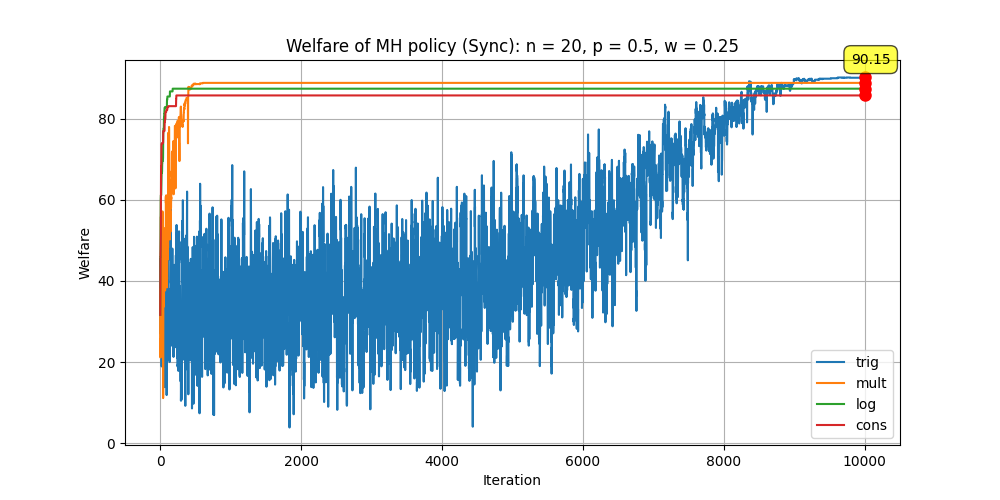}
		\caption{MH-policy: $n = 20$, $p = 0.5$, $\omega = 0.25$}
		\label{fig: WO_sync_0.25}
	\end{subfigure}
	\hfill
	\begin{subfigure}{0.49\textwidth}
		\centering
		\includegraphics[width = \linewidth]{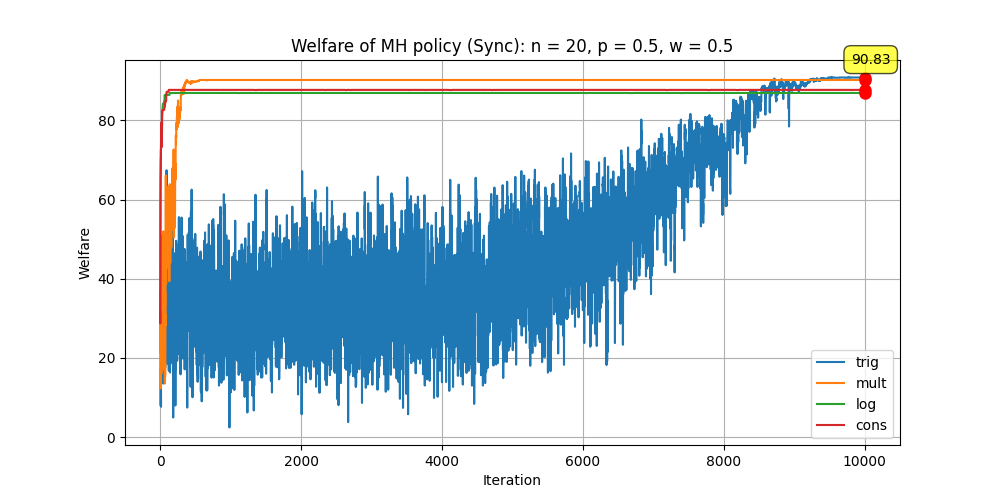}
		\caption{MH-policy: $n = 20$, $p = 0.5$, $\omega = 0.5$}
		\label{fig: WO_sync_0.5}
	\end{subfigure}
	
	\begin{subfigure}{0.49\textwidth}
		\centering
		\includegraphics[width = \linewidth]{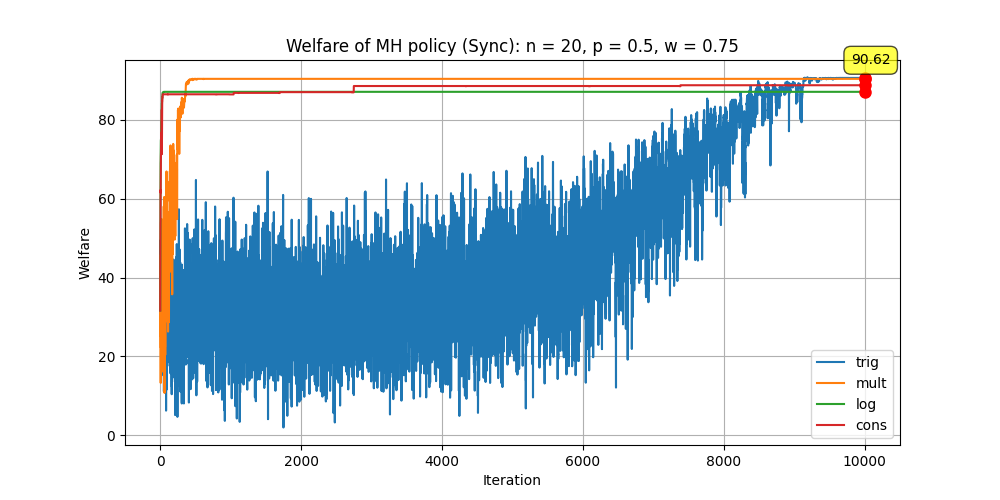}
		\caption{MH-policy: $n = 20$, $p = 0.5$, $\omega = 0.75$}
		\label{fig: WO_sync_0.75}
	\end{subfigure}
	\hfill
	\begin{subfigure}{0.49\textwidth}
		\centering
		\includegraphics[width = \linewidth]{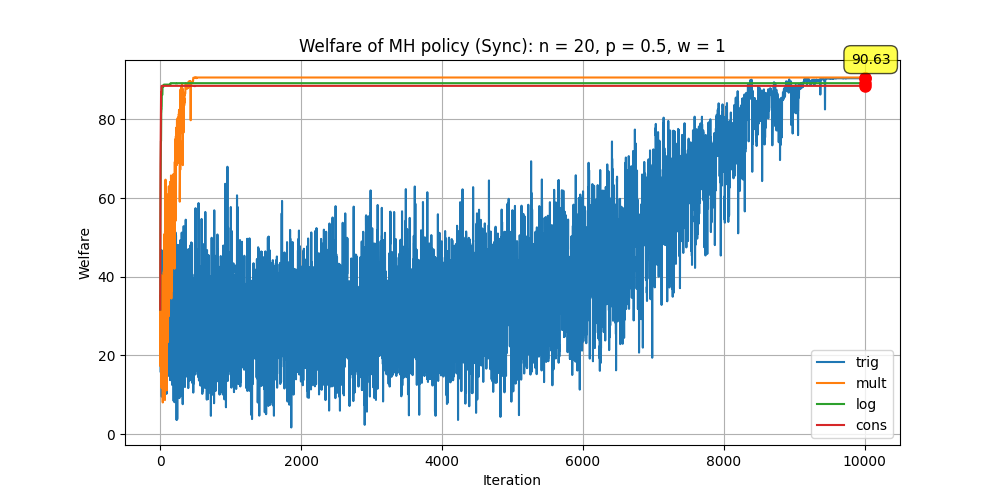}
		\caption{TS: $n = 20$, $p = 0.5$, $\omega = 1$}
		\label{fig: WO_sync_1}
	\end{subfigure}
	
	\caption{Dynamics of solving \ref{prob: optimalcoloring2} synchronously with different activation parameters.}
	\label{fig: WO_sync}
\end{figure}

In this group, we investigate on the performance of Algorithm~\ref{algo: MH-policy} under independently synchronous settings with different activation parameters $\omega$ (the game is completely synchronous when $\omega = 1$) while control the network features to be identical ($n = 20, p = 0.5$). One can compare the optimal welfare with that in \ref{fig: WO_async_TS_0.5} for effectiveness evaluation. See Figure~\ref{fig: WO_sync} for the results.

\subsection{Group D: RWO-Async}
\label{ssec: RWO-Async}
\begin{figure}[h]
	\centering
	\begin{subfigure}{0.49\textwidth}
		\centering
		\includegraphics[width = \linewidth]{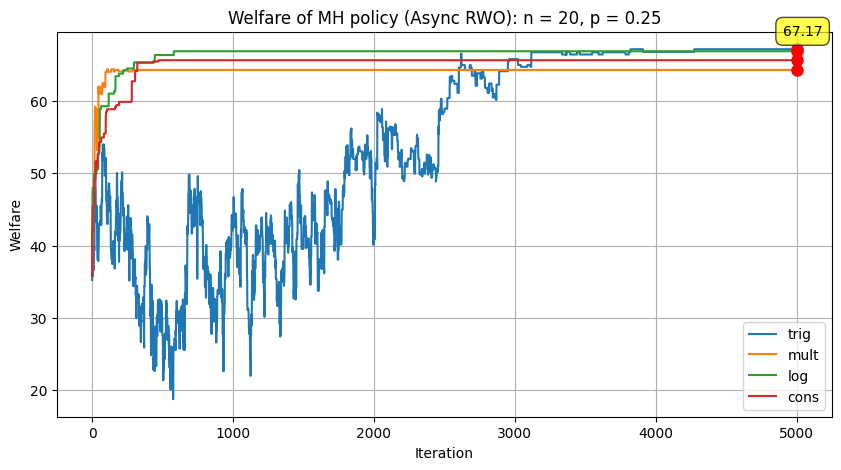}
		\caption{MH-policy: $n = 20$, $p = 0.25$}
		\label{fig: RWO_async_0.25}
	\end{subfigure}
	\hfill
	\begin{subfigure}{0.49\textwidth}
		\centering
		\includegraphics[width = \linewidth]{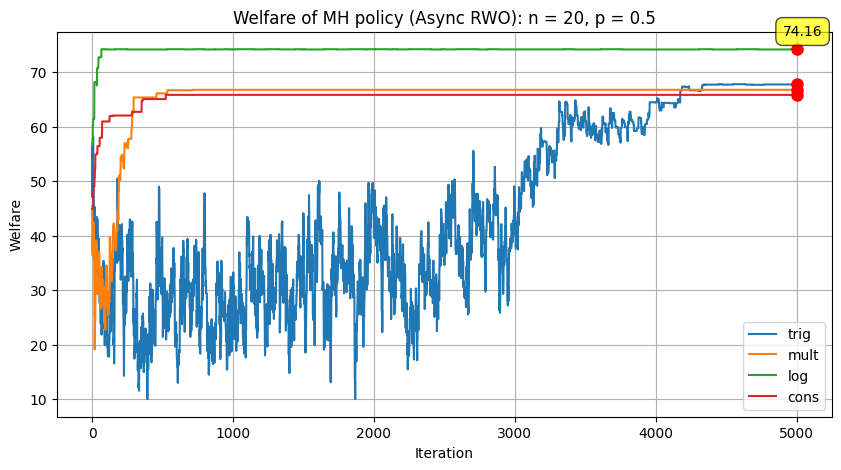}
		\caption{MH-policy: $n = 20$, $p = 0.5$}
		\label{fig: RWO_async_0.5}
	\end{subfigure}
	
	\begin{subfigure}{0.49\textwidth}
		\centering
		\includegraphics[width = \linewidth]{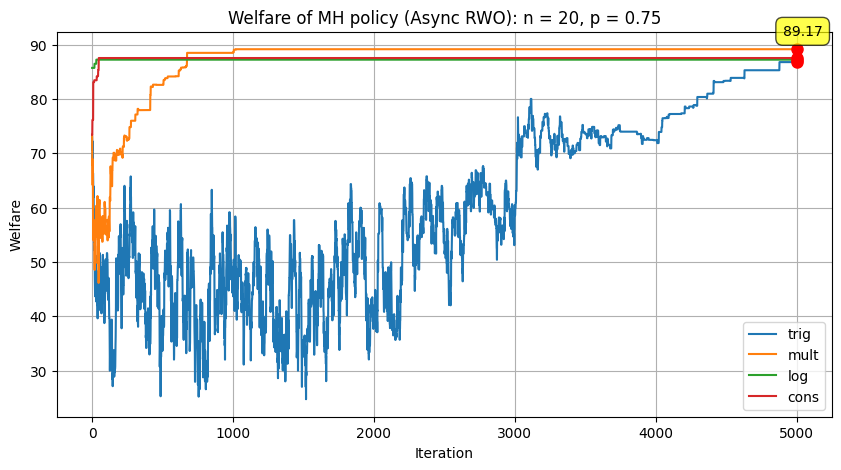}
		\caption{MH-policy: $n = 20$, $p = 0.75$}
		\label{fig: RWO_async_0.75}
	\end{subfigure}
	\hfill
	\begin{subfigure}{0.49\textwidth}
		\centering
		\includegraphics[width = \linewidth]{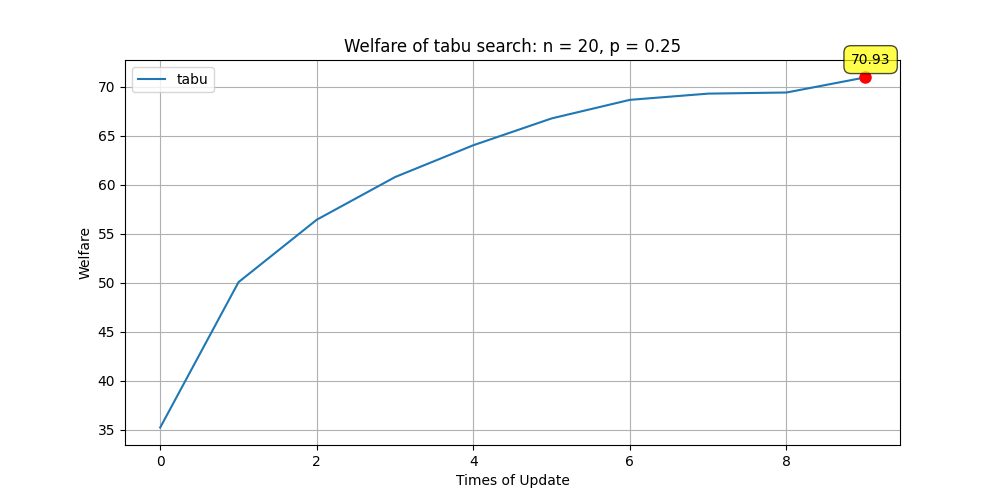}
		\caption{TS (for RWO): $n = 20$, $p = 0.25$}
		\label{fig: RWO_async_TS_0.25}
	\end{subfigure}
	
	\begin{subfigure}{0.49\textwidth}
		\centering
		\includegraphics[width = \linewidth]{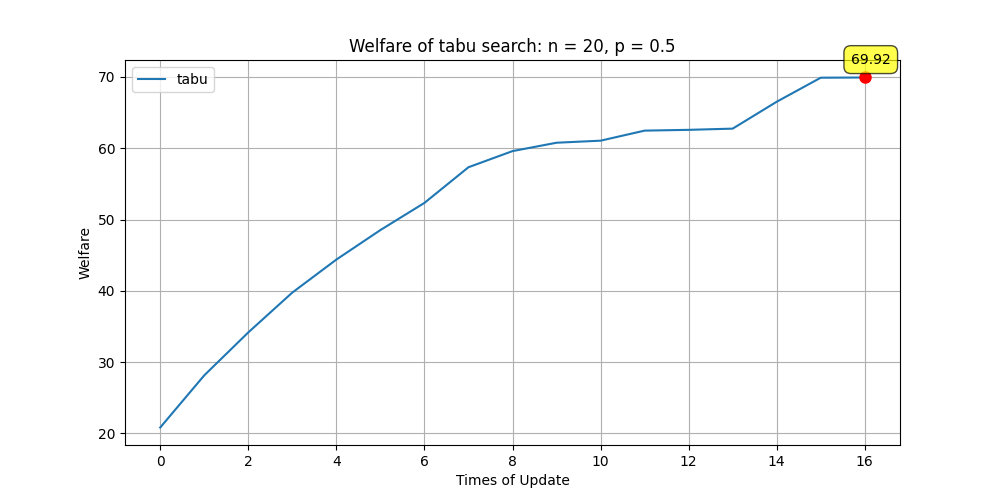}
		\caption{TS (for RWO):  $n = 20$, $p = 0.5$}
		\label{fig: RWO_async_TS_0.5}
	\end{subfigure}
	\hfill
	\begin{subfigure}{0.49\textwidth}
		\centering
		\includegraphics[width = \linewidth]{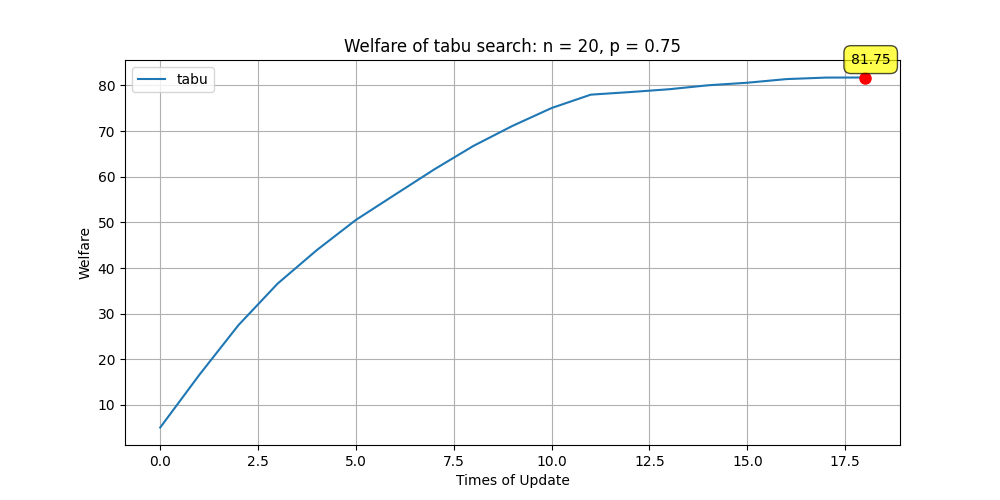}
		\caption{TS (for RWO): $n = 20$, $p = 0.75$}
		\label{fig: RWO_async_TS_0.75}
	\end{subfigure}
	
	\caption{Dynamics of solving \ref{prob: optimalcoloring2} for different network degrees.}
	\label{fig: RWO_async}
\end{figure}

In this group, we focus on Algorithm~\ref{algo: MH-policy_RWO} on solving \ref{prob: RWO}. The expected loss term highly depends on the conncection probabilities of the complementary edges which are indirectly determined by the network degrees. Therefore, we investigate on the asynchronous performance under networks with a same number of vertices ($n = 20$) yet different priori connection probabilities, as in Group B. See Figure~\ref{fig: RWO_async} for the results.

\subsection{Group E: RWO-Sync}
\label{ssec: RWO-Sync}
\begin{figure}[h]
	\centering
	\begin{subfigure}{0.49\textwidth}
		\centering
		\includegraphics[width = \linewidth]{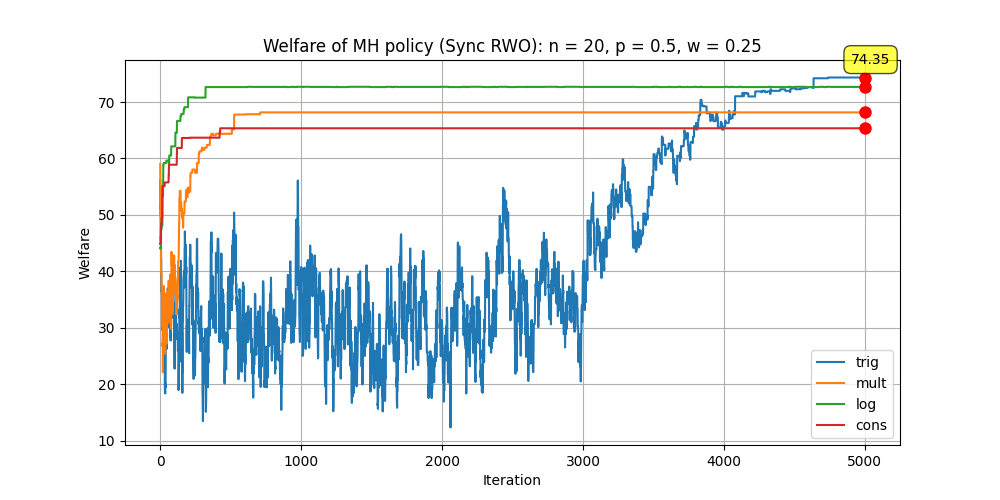}
		\caption{MH-policy: $n = 20$, $p = 0.5$, $\omega = 0.25$}
		\label{fig: RWO_sync_0.25}
	\end{subfigure}
	\hfill
	\begin{subfigure}{0.49\textwidth}
		\centering
		\includegraphics[width = \linewidth]{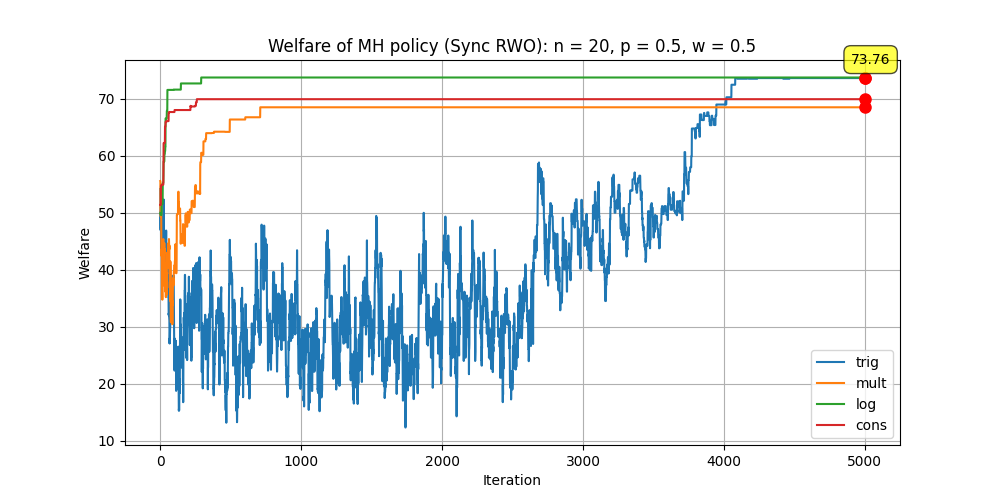}
		\caption{MH-policy: $n = 20$, $p = 0.5$, $\omega = 0.5$}
		\label{fig: RWO_sync_0.5}
	\end{subfigure}
	
	\begin{subfigure}{0.49\textwidth}
		\centering
		\includegraphics[width = \linewidth]{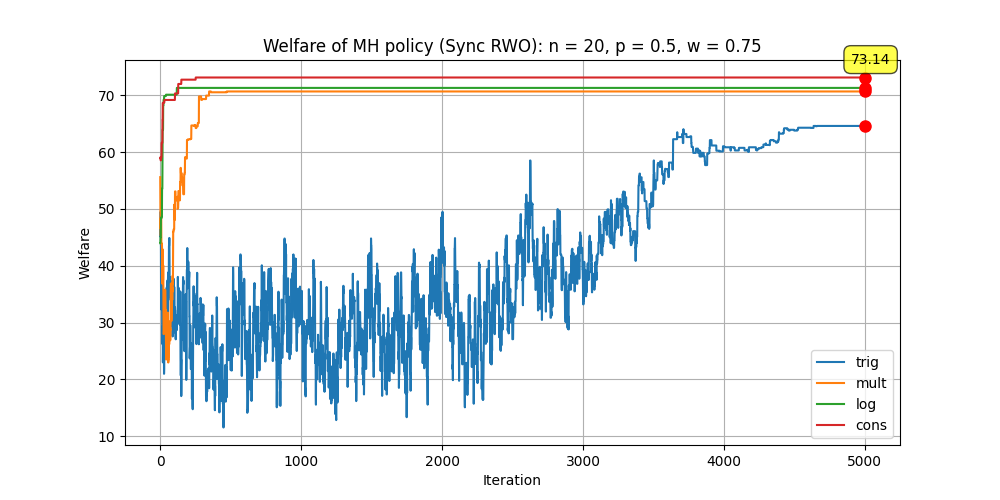}
		\caption{MH-policy: $n = 20$, $p = 0.5$, $\omega = 0.75$}
		\label{fig: RWO_sync_0.75}
	\end{subfigure}
	\hfill
	\begin{subfigure}{0.49\textwidth}
		\centering
		\includegraphics[width = \linewidth]{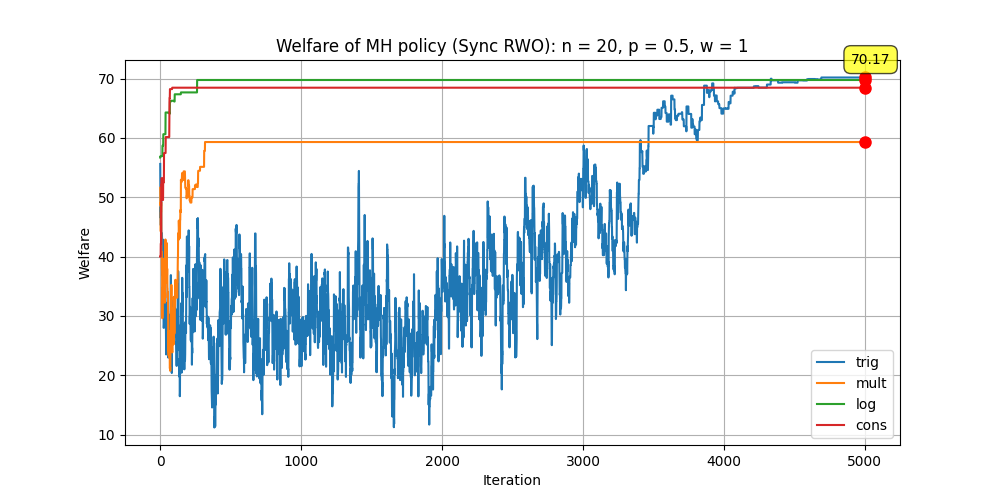}
		\caption{TS: $n = 20$, $p = 0.5$, $\omega = 1$}
		\label{fig: RWO_sync_1}
	\end{subfigure}
	
	\caption{Dynamics of solving \ref{prob: optimalcoloring2} synchronously with different activation parameters.}
	\label{fig: RWO_sync}
\end{figure}

In this group, we again investigate on the influence made by different activation parameters under synchronous settings as in Group C, yet focus on solving \ref{prob: RWO}. One can compare the optimal welfare with that in \ref{fig: RWO_async_TS_0.5}. See Figure~\ref{fig: RWO_sync} for the results.

\end{document}